\documentclass[12pt]{iopart}
\usepackage{graphicx}
\usepackage{dcolumn}
\usepackage{bm}
\usepackage{orcidlink}
\usepackage{subcaption}
\usepackage{gensymb}
\usepackage{tabularx}
\usepackage{float}
\begin{document}

\title[Gravitational Wave Peeps from EMRIs]{Gravitational Wave Peeps from EMRIs and their Implication for LISA Signal Confusion Noise}%

\author{Daniel J Oliver \orcidlink{0000-0002-7374-6925}$^{1,2}$, Aaron D Johnson \orcidlink{0000-0002-7445-8423}$^{1,3,4}$, Joel Berrier$^5$, Kostas Glampedakis \orcidlink{0000-0003-1860-0373}$^6$, Daniel Kennefick \orcidlink{0000-0002-5219-456X}$^{1,2}$}

\address{$^1$ Department of Physics, University of Arkansas, Fayetteville, AR 72701, USA}
\address{$^2$ Arkansas Center for Space and Planetary Sciences, University of Arkansas, Fayetteville, AR 72701, USA}%
\address{$^3$ Center for Gravitation, Cosmology and Astrophysics, University of Wisconsin-Milwaukee, Milwaukee, WI 53201, USA}%
\address{$^4$ Theoretical Astrophysics Group, California Institute of Technology, Pasadena, CA 91125, USA}
\address{$^5$ Department of Physics, Astronomy, University of Nebraska-Kearney, Kearney, NE 68849, USA}%
\address{$^6$ Departamento de Física, Universidad de Murcia, Murcia E-30100, Spain}
\ead{mail@danieljoliver.com}
\vspace{10pt}

\begin{abstract}
Scattering events around the center of massive galaxies will occasionally toss a stellar-mass compact object into an orbit around the massive black hole at the center, beginning an extreme mass ratio inspiral. The early stages of such a highly eccentric orbit are not likely to produce detectable gravitational waves, as the source will only be in a suitable frequency band briefly when it is close to periapsis during each long-period orbit. This repeated burst of emission, firmly in the millihertz band, is the gravitational wave peep. While a single peep is not likely to be detectable, if we consider an ensemble of such subthreshold sources, spread across the universe, together they may produce an unresolvable background noise that could obscure sources otherwise detectable by the Laser Interferometer Space Antenna. Previous studies of the extreme mass ratio signal confusion background focused either on parabolic orbits near the massive black hole or events closer to merger. We seek to improve this characterization by implementing numerical kludge waveforms that can calculate highly eccentric orbits with relativistic effects. Our focus is on orbits at the point of capture that are farther away from the massive black hole. Here we present the waveforms and spectra of peeps generated from recent calculations of extreme mass ratio inspirals/bursts capture parameters and discuss how these can be used to estimate the signal confusion noise generated by such events. We demonstrate the effects of changing the orbital parameters on the resulting spectra as well as showing direct comparisons to parabolic orbits and why the gravitational wave ``peep" needs to be studied further. The results of this study will be expanded upon in a further paper that aims to provide an update on the EMRI signal confusion noise problem.
\end{abstract}


\maketitle

\section{Introduction}

Gravitational waves were first detected with the Laser Interferometer Gravitational Wave Observatory (LIGO) on September 14th, 2015, generated from two black holes on the order of $\sim30\ M_{\odot}$ \cite{abbott_firstdetection_2016}. While LIGO allowed us access to high-frequency gravitational wave sources at low redshift, the Laser Interferometer Space Antenna (LISA), the proposed ESA-NASA space-based gravitational wave detector, will allow us to look into lower frequencies ($0.1$ mHz - $0.1$ Hz) and expand our range to redshifts of $z \sim 20$. \cite{amaro-seoane_laser_2017,babak_kludge_2007,barack_confusion_2004,barack_lisa_2004,bonetti_gravitational_2020}. The technology behind LISA has been successfully tested by LISA Pathfinder which produced outstanding results showing that the drag-free technology can work with great accuracy \cite{babak_science_2017}.\par

Observations suggest that almost all galaxies contain at their centers, a massive black hole (MBH), with a mass range of $10^4\ M_\odot$ – $10^9\ M_\odot$. This includes our own Milky Way Galaxy, whose central black hole has a mass of $\sim 4.31 \times 10^6\ M_\odot$ \cite{bonetti_gravitational_2020,kormendy_inward_1995,kormendy_coevolution_2013,reynolds_observational_2021}. These are believed to be massive Kerr (spinning) black holes \cite{amaro-seoane_laser_2017,bonetti_gravitational_2020,glampedakis_zoom_2002}. The inner few parsecs around the MBH contain a dense cusp of millions of stars, including significant numbers of stellar-mass compact objects (COs) such as black holes, neutron stars, or white dwarfs. Two-body gravitational interactions will occasionally toss one of those bodies into a highly eccentric orbit around the MBH. These events give rise to Extreme Mass Ratio Inspirals (EMRIs) and are expected to be a major source of gravitational wave detections for LISA. The mass ratio is defined by $q\equiv\mu/M$ , where $\mu$ is the CO mass and $M$ is the MBH mass. EMRI orbits typically have a very small mass ratio $q \sim 10^{-7}-10^{-4}$ and can be modeled using black hole perturbation theory \cite{babak_kludge_2007,babak_science_2017,barack_confusion_2004,barack_lisa_2004,bonetti_gravitational_2020,glampedakis_zoom_2002}. Due to the extreme mass ratio, the orbital parameters will evolve adiabatically with radiation reaction, meaning that the parameters will evolve on a much longer time scale than the orbital period. These are typically highly eccentric long-period orbits at the point of capture, and because they evolve adiabatically we can assume that the EMRI is in a state of approximate equilibrium such that the orbital parameters do not decay over a single orbit \cite{amaro-seoane_laser_2017,babak_kludge_2007,babak_science_2017,barack_confusion_2004,barack_lisa_2004,bonetti_gravitational_2020,gair_prospects_2017,glampedakis_zoom_2002}.\par

There are two types of capture events for compact objects falling into MBHs: the first is a direct plunge event and the second is a slow inspiral which is the type that will be focused on in this paper. These types of orbits can be modeled by treating the CO as a test particle moving in the gravitational field of the MBH. At the point of capture, these are considered generic orbits (but with eccentricity $e \approx 1$) with an inclination with respect to the MBH's equatorial plane. As the CO passes the pericenter on its highly eccentric orbit it will emit gravitational radiation carrying away energy and angular momentum which will then cause the orbit to evolve secularly, circularizing and inspiraling closer to the central black hole forming the familiar EMRI orbit \cite{babak_kludge_2007,babak_science_2017,barack_lisa_2004, yunesRelativisticEffectsExtreme2008, glampedakis_approximating_2002}. \par

Eccentric EMRI orbits tend to have orbits that exhibit ``Zoom-Whirl” behavior (these orbits are also called homoclinic orbits \cite{glampedakis_approximating_2002,levinGravityWavesHomoclinic2000}). These are characterized by a relatively long period gravitational wave feature associated with the complete orbit, but with a much shorter period feature at periapsis. It is normal, of course, in Keplerian orbits, for the orbiting body to zoom into periapsis at its highest orbital velocity (Kepler’s second law), but relativistic effects associated with strong spacetime curvature and (sometimes) gravitomagnetic effects produce a complete cycle, sometimes several complete cycles, of this higher frequency emission at periapsis. This latter feature is the whirl. Zoom-whirl orbits near the end of the inspiral are of interest as suitable candidates for detection by LISA. It has also been observed that parabolic orbits can produce whirl-like behavior (though typically with only one whirl) as they swing (or ``zoom”) by periapsis. These are known as extreme mass ratio bursts (EMRBs), since they do not form part of a continuous or periodic gravitational wave signal. In between these two extremes, there exist highly eccentric orbits, most probably not individually detectable, which nevertheless produce ``whirl” or ``burst” like features at periapsis repeatedly throughout their history \cite{glampedakis_approximating_2002,rubboEventRateExtreme2006, hopmanGravitationalWaveBursts2007,yunesRelativisticEffectsExtreme2008,toonenGravitationalWaveBackground2009, berryObservingGalaxyMassive2013a, berryExtrememassratioburstsExtragalacticSources2013a, berryExpectationsExtrememassratioBursts2013a,   fanExtrememassratioBurstDetection2022}. In order to draw attention to the periodic, repetitive nature of these features, we will refer to them as ``peeps\footnote{The short high-pitched signal coming amidst a low-frequency background suggests a bird sound, but the word "chirp" is already in use in this field.}” in this paper. Each EMRI is to be thought of as a small bird, occasionally peeping. The ensemble of EMRIs in the nearby Universe creates a background of peeps. In addition, the peep evolves in ``chirp” fashion during the inspiral, both in frequency and amplitude. Indeed, in amplitude terms, the peep is the most prominent feature of each EMRI chirp throughout most of its evolution. Unfortunately, the ensemble of ``peeps" from around the Universe is likely to constitute a signal confusion noise which may be an issue for LISA signal analysis. Hence the purpose of this paper is to study the ``peep" waveform and spectrum more closely.\par

\noindent Throughout this paper we use geometric units $G=c=1$.

\section{Background}

\subsection{Semi-relativistic Approximation Waveforms}

To model waveforms from early-stage EMRIs, we used a time domain code developed by Babak et al (2007). The “numerical kludge” (NK) produces an inspiral trajectory in phase space defined by orbital energy (E), axial angular momentum (L), and the Carter constant (Q). It then numerically integrates the Kerr geodesic equations along the inspiral to obtain the Boyer-Lindquist coordinates $\{t,r,\theta,\phi\}$ of the test particle. With these coordinates, it constructs the gravitational waveform from the inspiral trajectory \cite{babak_kludge_2007}\footnote{For a more detailed description of the methods behind the numerical kludge see Babak et al (2007) which discusses the full calculations used in the code \cite{babak_kludge_2007}.}.\par 

The NK waveforms utilize a semi-relativistic approximation (first developed by Ruffini et al 1981 \cite{ruffini_semi-relativistic_1981}) with initial parameters of MBH spin ($a/M$), semi-latus rectum ($p/M$), orbital eccentricity ($e$), orbital inclination ($\iota$), the mass ratio ($q$), as well as the viewing angles $\{\Theta, \Phi \}$ (The latitude and azimuthal angles respectively). To better show the effects due to changing the various parameters, in this paper we will utilize both gravitational wave polarizations $h_+$ and $_\times$. \par
The semi-relativistic approximation works very well for $r_p>5M$ (which for our high eccentricities $e\simeq 1$ is equivalent to $p>10M$ using $r_p=\frac{p}{1+e}$) when compared to more exact methods such as a Teukolsky based approach \cite{babak_kludge_2007, gair_semi-relativistic_2006}. The numerical kludge captures more relativistic effects at the cost of computation time compared to the analytical kludge model (AK). The augmented analytical kludge (AAK) model was not implemented as it is unable to model the EMRI waveforms at the eccentricity needed for this project (See Figs. 1 and 2 in Isoyama (2022) which shows the errors for eccentricity up to $e=0.4$) \cite{chua_augmented_2017, isoyama_2022}. FastEMRIWaveforms (FEW) was not implemented as it can currently only calculate the Schwarzschild equatorial case where $a=0$ and $\iota=0^\circ$ \cite{FEW1,FEW2}.

\subsection{Extreme Mass Ratio Bursts (EMRBs)}

Extreme mass ratio bursts occur when a stellar-mass compact object encountering a massive black hole has a pericenter passage with a timescale of less than $10^5\ s$. They are traditionally modeled as parabolic orbits with $e=1$ \cite{rubboEventRateExtreme2006}. However bursts with eccentricity close to unity are the precursors to EMRIs which are a periodically repeated burst or ``peep," with a very long lapse between peeps. On these fly-by orbits, the CO passes through periapsis, where there is a beamed burst of gravitational radiation which carries away energy and angular momentum from the system. This emission of gravitational waves causes the orbital period to decrease and leads to the orbit circularizing over time forming an extreme mass ratio inspiral \cite{rubboEventRateExtreme2006, hopmanGravitationalWaveBursts2007,yunesRelativisticEffectsExtreme2008,toonenGravitationalWaveBackground2009, berryObservingGalaxyMassive2013a, berryExtrememassratioburstsExtragalacticSources2013a, berryExpectationsExtrememassratioBursts2013a,   fanExtrememassratioBurstDetection2022}. \par

Previous studies of these bursts have discussed the detectable event rates for eLISA \cite{berryExpectationsExtrememassratioBursts2013a,berryExtrememassratioburstsExtragalacticSources2013a,berryObservingGalaxyMassive2013a}, LISA \cite{rubboEventRateExtreme2006, hopmanGravitationalWaveBursts2007,toonenGravitationalWaveBackground2009}, and TianQin \cite{fanExtrememassratioBurstDetection2022}. The rates seem to be on the order of $\sim 1\ yr^{-1}$ for the Milky Way Galaxy and lower for extra-galactic sources. These extra-galactic signals require a very close approach of the CO to the MBH for the amplitude to be large enough for detection. \par
Much of the previous work on EMRBs focused primarily on orbits that brought the CO close to the MBH as these are likely to be detectable with parameter estimation being possible. The waveform calculations for these, however, were utilizing either Newtonian waveforms or the numerical kludge. Both of these dramatically lose accuracy when the orbiting particle is close to the MBH \cite{babak_kludge_2007,rubboEventRateExtreme2006, hopmanGravitationalWaveBursts2007,yunesRelativisticEffectsExtreme2008,toonenGravitationalWaveBackground2009, berryObservingGalaxyMassive2013a, berryExtrememassratioburstsExtragalacticSources2013a, berryExpectationsExtrememassratioBursts2013a,   fanExtrememassratioBurstDetection2022}.

\subsection{Background Signal Confusion Noise}

 In the very late stages of the inspiral, EMRIs are expected to be a major source of gravitational wave detections by LISA. However, during the early stages of the inspiral, the ``peep" part of the signal recurs so rarely as to likely be undetectable. The signal-to-noise ratio from a single peep is generally below the detectability level of LISA since a single peep is too brief and low in amplitude to be seen. However, an ensemble of these sources will create a potentially unresolvable background noise \cite{barack_confusion_2004,bonetti_gravitational_2020,racine_gaussianity_2007}.\par

Previous studies of this EMRI signal confusion noise were undertaken by \cite{barack_confusion_2004, bonetti_gravitational_2020, chua_non-local_2022, racine_gaussianity_2007}. These studies primarily focused on the EMRI waveforms when they are emitting gravitational waves in LISA's frequency band much more frequently. Barack and Cutler (2004) utilized approximated Schwarzschild (non-rotating) EMRI spectra generated from the analytical kludge model. Bonetti and Sesana (2020) utilized a similar waveform model while implementing updated relativistic population data from Babak et al. (2017) as well as a more recent LISA sensitivity curve \cite{babak_science_2017, barack_confusion_2004, bonetti_gravitational_2020}. \par 

Some additional studies on the effects of EMRB sources on the gravitational wave background have also been performed \cite{toonenGravitationalWaveBackground2009,fanExtrememassratioBurstDetection2022}. Toonen et al. (2009), observed the potential gravitational wave background from such EMRB sources for LISA using a Newtonian order approximation for waveforms and found that the EMRB background constituted a signal that was an order of magnitude below the LISA instrumental noise \cite{toonenGravitationalWaveBackground2009}. They also found that by using the O'Leary et al. (2009) model the EMRB background was smaller than the LISA instrumental noise by a factor of $\sim 1.5$ \cite{olearyGravitationalWavesScattering2009}.
Fan et al. (2022), took the sources which were considered not detectable and analyzed the gravitational wave background generated from a single pass of these sources during one year of observation time for TianQin and found that the background was 6 orders of magnitude lower than the detectable range \cite{fanExtrememassratioBurstDetection2022}.\par

Our study begins with determining the estimated capture parameters of EMRIs based on updated literature and then modeling highly eccentric EMRIs from the point of capture. We focus on the evolution of the peep during an inspiral as a first step in understanding its ability to mask sources that would otherwise be observable.\par 
We will provide waveforms and spectra of EMRIs with peeps, paying attention to variations of peeps depending on EMRI parameters, such as the spin of the MBH, represented by $a/M$, the eccentricity of the orbit denoted by $e$, the semi-latus rectum as $p/M$, and the orbital inclination $\iota$ (see Table \ref{tab:Param}).

\subsection{Model Parameters}

\begin{table}[h!]
\caption{\label{tab:Param}EMRI Parameters at Capture from Literature.}
\begin{center}
\item[]\begin{tabular}{@{}llll}
\br
Parameter & Value & Citation\\
\mr
Black Hole Spin (a/M) & $0.7-0.95$ & Gammie et al (2004) \cite{gammie_black_2004}\\ 
& $0.66-0.99$ & Reynolds (2021)\cite{reynolds_observational_2021}\\ \cline{1-3}
Semi-Latus Rectum (p/M)   & $15-120$ & Freitag (2003)\cite{freitag_gravitational_2003} \\\cline{1-3}
Eccentricity (e)  & 0.999-0.999999 & Freitag (2003)\cite{freitag_gravitational_2003}  \\
& 0.99-0.999999 & V\'{a}zquez-Aceves et al (2022)\cite{vazquez-aceves_revised_2022} \\\cline{1-3}
Orbital Inclination ($\iota$) &  $0^\circ-180^\circ$\\
\br
\end{tabular}
\end{center}
\end{table}

Barack and Cutler (2004) modeled signal confusion noise with the orbits of EMRIs farther into their lifespan, specifically near plunge when the EMRIs would be detectable by LISA. They state the parameters of EMRIs at capture as they were known at the time.\par

The parameters at capture for the radii at periapsis were found to be $r_p = 8M-100M$, with an orbital eccentricity  of $0.999<e<0.999999$. These were determined by Freitag (2003) who performed a Monte Carlo simulation studying the stellar mass objects in the cusp of the Milky Way galaxy scattered due to gravitational interactions \cite{freitag_gravitational_2003}. The radius at periapsis values were then converted into semi-latus rectum using $r_p=\frac{p}{1+e}$ with the eccentricity range from the same source. The MBH spins were found to be between $0.7M<a<0.95M$ by Gammie et al. (2004) \cite{gammie_black_2004}.\par

Some recent work has been done to update the parameters of EMRI orbits and MBH, specifically the capture eccentricity and the MBH spin. Reynolds (2021) implemented X-ray reflection spectroscopy and thermal continuum fitting to measure the spins of accreting MBHs. MBHs were found to be rapidly spinning with an average spin of  $a/M \sim 0.9$ for masses $M<3\times10^{7}M_{\odot}$. For larger masses, Reynolds (2021) found moderately spinning MBHs with an average spin of $a/M \sim 0.8$ (summarized in Table \ref{tab:Reyn}) \cite{reynolds_observational_2021}. \par

\begin{table}
\caption{\label{tab:Reyn}Summary of published MBH/AGN spin measurements using X-ray reflection method from Reynolds (2021), separated into two ranges of masses. In Reynolds' paper it was stated that there were higher average spins until $M = 3\times10^7M_\odot$\cite{reynolds_observational_2021}.}
\begin{center}
\item[]\begin{tabular}{@{}llll}
\br
Mass Range  & Spin Range  & Spin Average \\
$(\times10^6 M_\odot$) & $(a/M)$ & $(a/M)$\\
\mr
$\sim1.1-29.8^{+5.4}_{-5.4}$ & $0.66^{+0.30}_{-0.54}-0.99$ & $\sim0.912$\\
$\sim34-4500^{+1500}_{-1500}$ & $0.4-0.98$ & $\sim0.782$\\
\br
\end{tabular}
\end{center}
\end{table}

Previous studies of EMRI signal confusion noise focused primarily on equatorial orbits (inclination $\iota=0$). To better encapsulate the problem, our study is expanding to generic orbits where we will be exploring the full inclination parameter space from $0^\circ$ to $180^\circ$. This will allow us to look at prograde and retrograde orbits. \par

To our knowledge, no studies were found to improve upon the initial work for the semi-latus rectum at capture determined by Freitag (2003). However, after re-examining the data from Freitag's Monte Carlo simulation, we chose to focus only on the higher concentrated regions of the simulated ``stars" at capture, thereby constraining the radius at periapsis to $r_p=30M$. More work has been done for the eccentricity at capture with recent estimates of $0.99<e<0.999999$ found by V\'{a}zquez-Aceves et al (2022) \cite{vazquez-aceves_revised_2022}.\par

A compilation of our estimated parameters at capture useful to our study is shown in Table \ref{tab:KParam}. These encompass higher spin MBH as shown from \cite{reynolds_observational_2021}, slightly lower semi-latus rectum values converted from \cite{freitag_gravitational_2003}, eccentricity ranges from \cite{freitag_gravitational_2003,vazquez-aceves_revised_2022}, and finally inclined orbits which will cover the range of $0^\circ\leq\iota\leq180^\circ$.\par

\begin{table}
\caption{\label{tab:KParam}EMRI Capture Parameters useful for gravitational wave ``peeps".}
\begin{center}
\item[]\begin{tabular}{@{}llll}
\br
Parameter & Minimum Value & Maximum Value\\
\mr
MBH Spin, $a$ & 0.8M & 0.9M\\ \hline
Semi-Latus Rectum, $p$ & 15M & 120M\\ \hline
Eccentricity, $e$ & 0.99 & 0.999999 \\ \hline 
Inclination, $\iota$ & $0^{\circ}$ & $180^{\circ}$\\
\br
\end{tabular}
\end{center}
\end{table}

\section{Results}
\subsection{Peep Waveforms}

\begin{figure}[hb!]
  \begin{subfigure}[b]{1\linewidth}
    \centering
    \fbox{\includegraphics[width=0.75\linewidth]{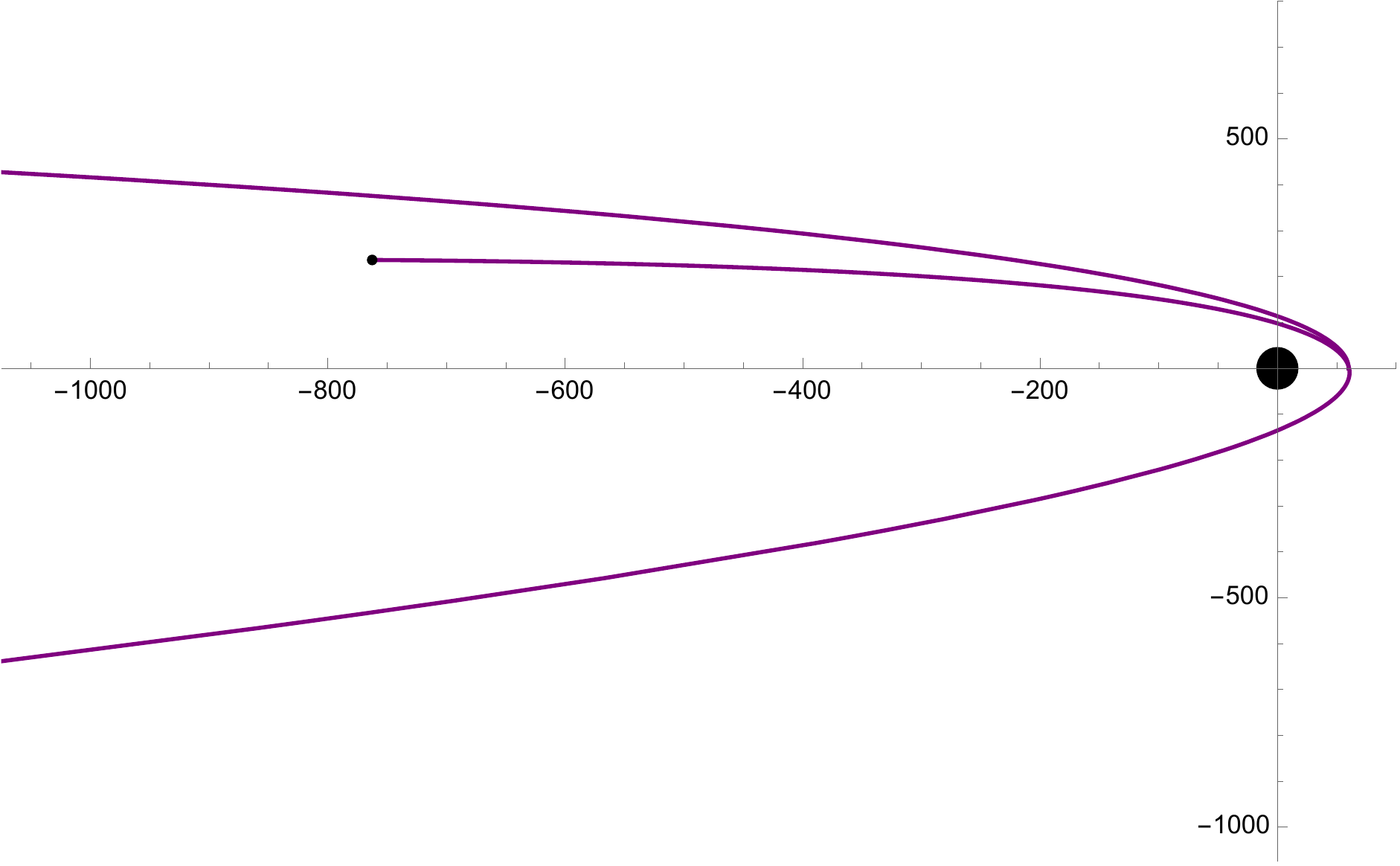}}
  \end{subfigure} 
  \caption{Kerr geodesic orbits created using the Black Hole Perturbation Toolkit \cite{BHPToolkit} for parameters of $a=0.9M$, $p=120M$, $e=0.999999$, $\iota=0.0^{\circ}$. The geodesic is plotted using Boyer-Lindquist coordinates with units of $M$ converted to  Cartesian coordinates. The figure zooms in on the region of the periapsis where the peep part of the gravitational waveform is produced. The orbit otherwise extends outward very far. This figure also shows periapsis precession as the CO exits the orbit at a different trajectory than it entered.}
  \label{fig:geo9}
\end{figure}

 To better visualize the orbit of the CO around the MBH, we show in Figure \ref{fig:geo9} the geodesic orbit as calculated using the Black Hole Perturbation Toolkit \cite{BHPToolkit}. From the geodesic orbit, one can see that the CO after completing one full orbit will have a slight phase shift as it leaves the proximity of the MBH. At its closest approach to the MBH the CO is quite close to the unstable circular orbit in the relativistic effective potential and thus emits a close-to-complete cycle gravitational wave at a frequency that is not far from that associated with that unstable circular orbit. The behavior is similar to the type of orbit (from later in the inspiral) known as a Zoom-Whirl orbit. The ``whirl", and in this case the ``peep" is associated with the prominent perihelion advance which can be clearly seen in Figure \ref{fig:geo9} \cite{glampedakis_zoom_2002}. While the single orbit itself is not likely to be detectable on its own, the frequency of the whirl part of the orbit could be detectable with a large enough number of whirls. Figure \ref{fig:geo9} shows an equatorial orbit with a spin of $a=0.9M$, a semi-latus rectum of $p=120M$, and an orbital eccentricity of $e=0.999999$.\par

\begin{figure*}
  \begin{subfigure}[b]{0.5\linewidth}
    \centering
    \includegraphics[width=\linewidth]{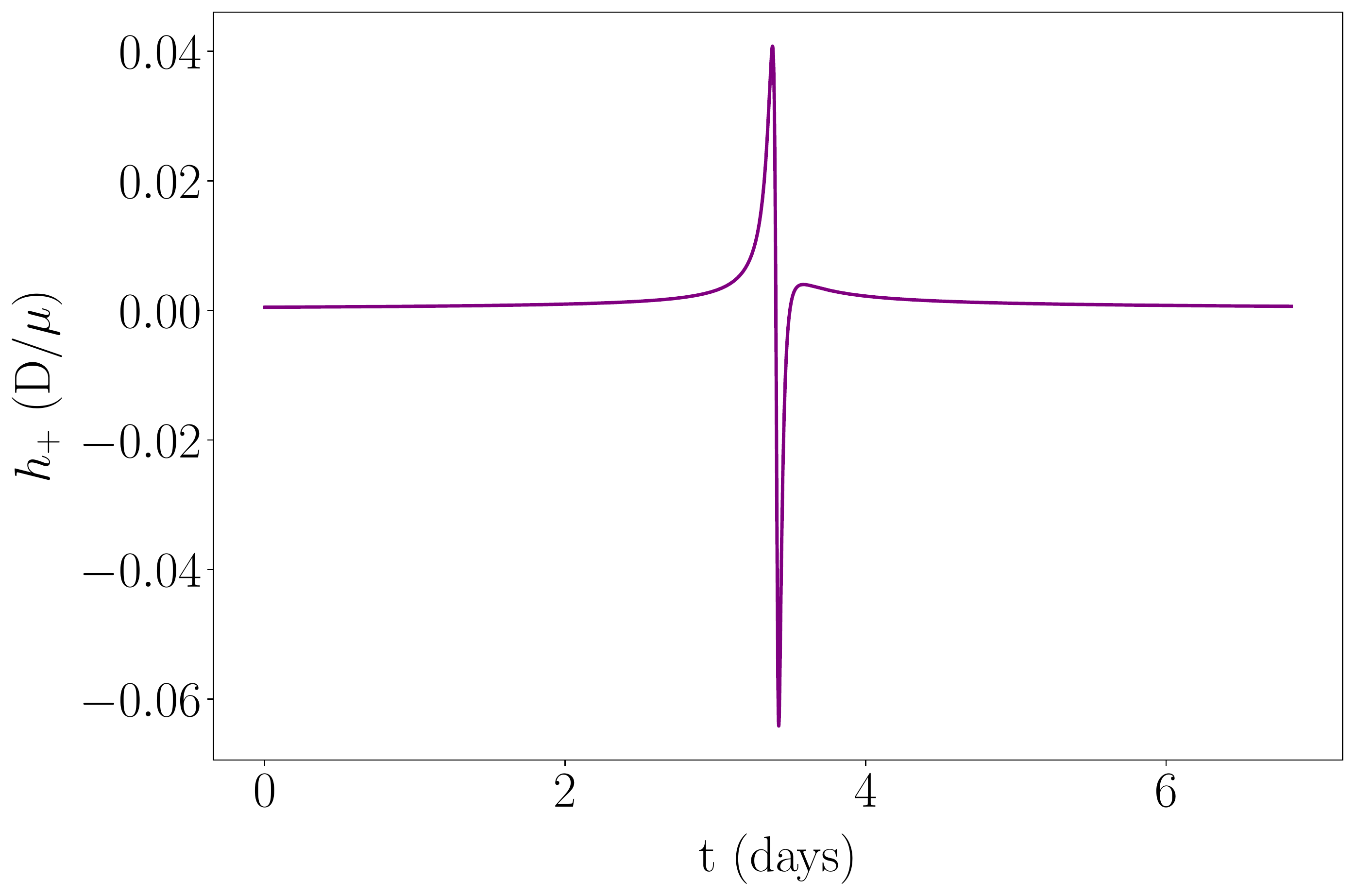} 
    \caption{$h_+$ Waveform}
  \end{subfigure}
  \begin{subfigure}[b]{0.5\linewidth}
    \centering
    \includegraphics[width=\linewidth]{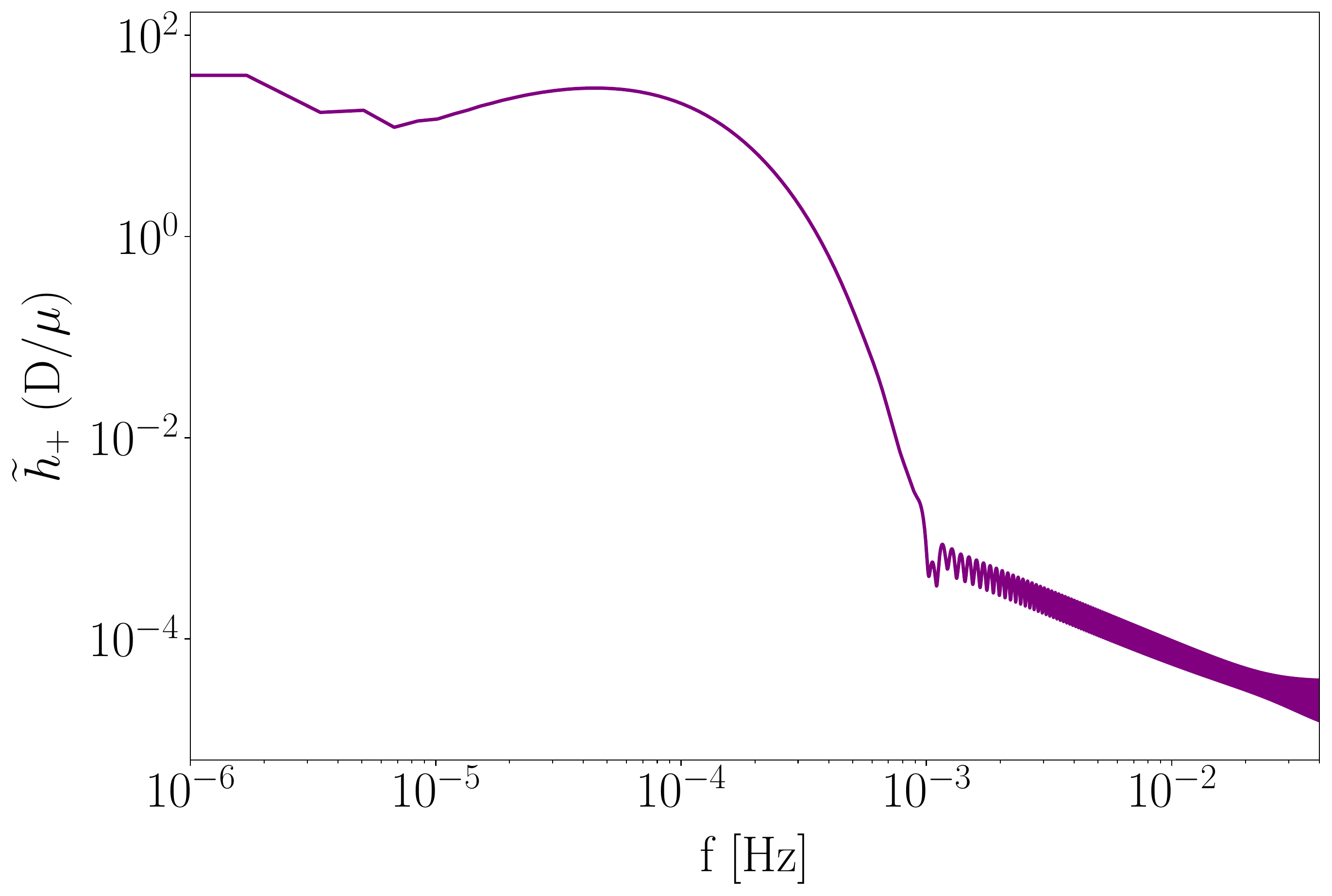}
    \caption{$h_+$ Spectra}
  \end{subfigure} 
  \begin{subfigure}[b]{0.5\linewidth}
    \centering
    \includegraphics[width=\linewidth]{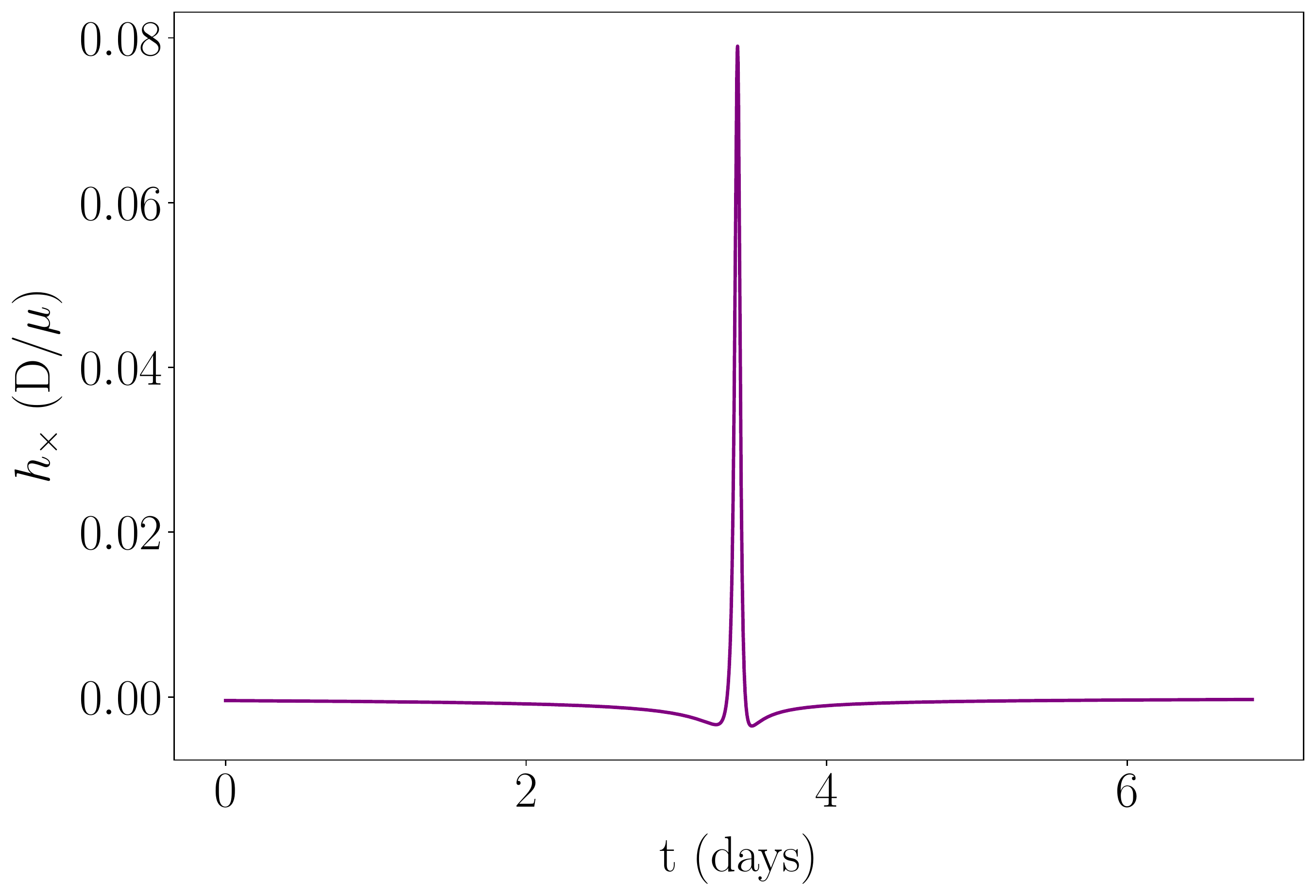} 
    \caption{$h_\times$ Waveform}
  \end{subfigure}
  \begin{subfigure}[b]{0.5\linewidth}
    \centering
    \includegraphics[width=\linewidth]{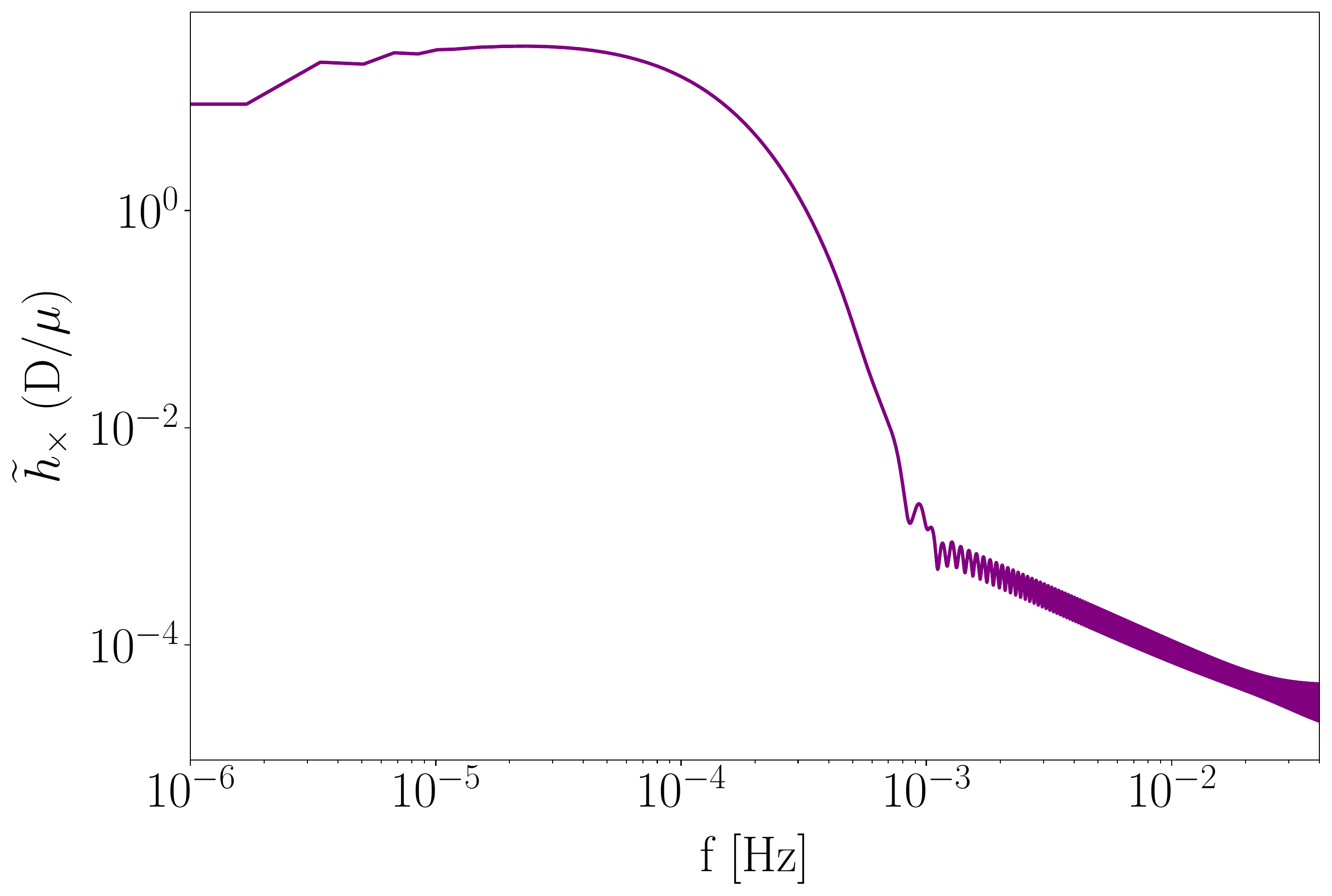}
    \caption{$h_\times$ Spectra}
  \end{subfigure} 
  \caption{NK peep waveform and spectra for $h_+$ and $h_\times$ created with $q=10^{-6}$, $a=0.9M$, $p=120M$, $e=0.999999$, $\iota=25.0^{\circ}$, $dt=10$ seconds, $\Theta=45^{\circ}$, and $\Phi=45^{\circ}$. The length of the data was chosen as $\approx10\times$ the width of the peep itself equivalent to $\approx7$ days of observation. The full orbital period at this point in the inspiral is $T\approx185000$ years. The waveforms shown in (a) and (c) have an x-axis as the time measured in days. The y-axis shows the $h_+$ (a) and the $h_\times$ (c) polarized waveforms scaled with $D$, the radial distance of the observational point, and $\mu$, the test body's mass. The spectra for the $h_+$ (b) and $h_\times$ (d) were created by performing an FFT of the corresponding waveform using a Tukey Window. The x-axis is frequency measured in Hertz. The y-axis is scaled by the same $D$ and $\mu$ as the waveform.}
  \label{fig:waveform}
\end{figure*}

The geodesic orbit can be further viewed as a time domain waveform such as in Figure \ref{fig:waveform} for both $h_+$ and $h_\times$ which shows a single period orbit at capture with parameters $q=10^{-6}$, $a=0.9M$, $p=120M$, $e=0.999999$, $\iota=25.0^{\circ}$, $dt=10$ seconds, $\Theta=0.0^{\circ}$, and $\Phi=90^{\circ}$.\par

\begin{figure*}[hb!]
  \begin{subfigure}[b]{1\linewidth}
    \centering
    \includegraphics[width=1\linewidth]{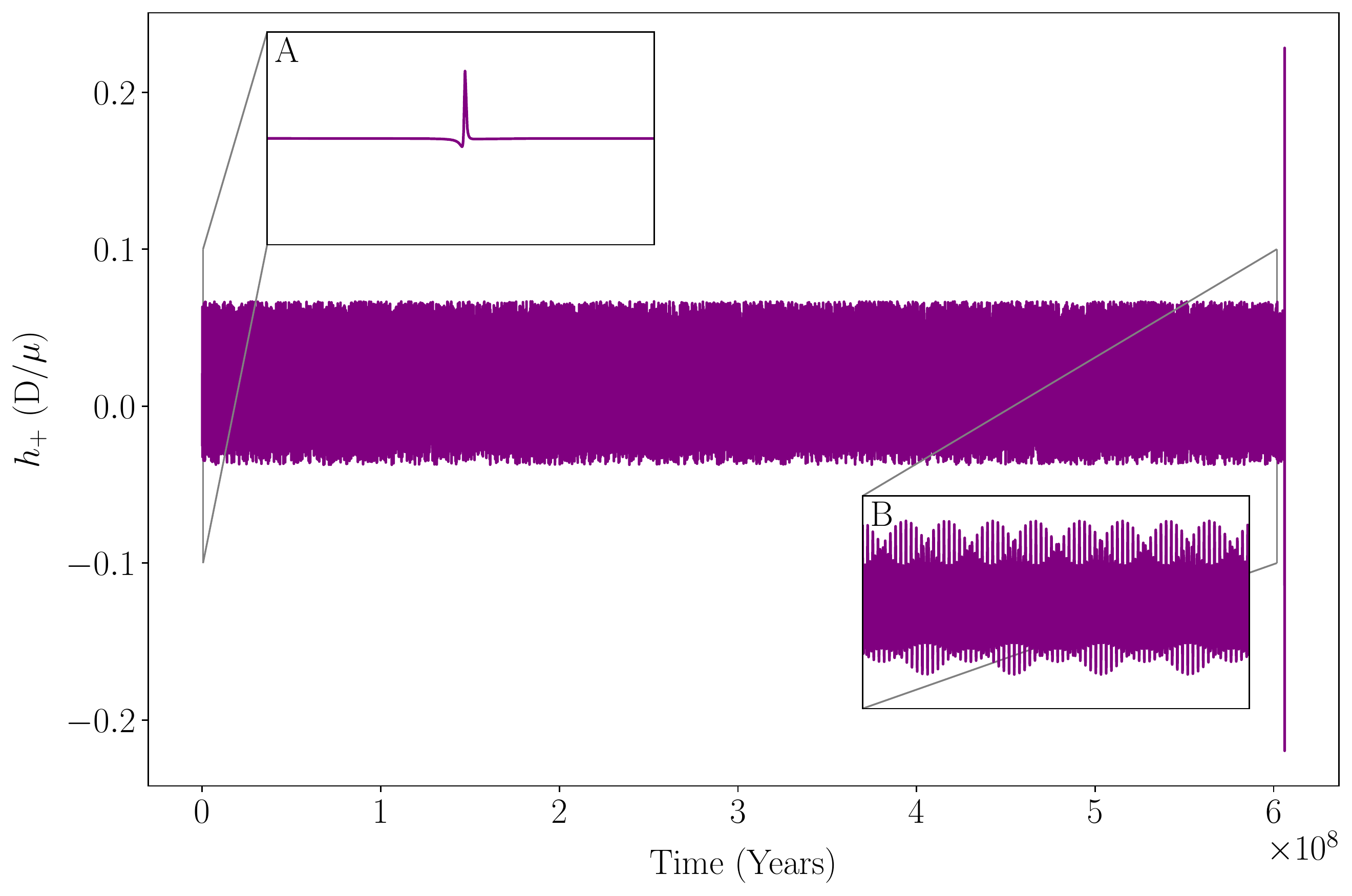} 
  \end{subfigure} 
  \caption{Fully evolved inspiral utilizing a mass ratio of $q=10^{-6}$ and initial parameters of $a=0.8M$, $p=119.999916M$, $e=0.9999981$, $\iota=0.0^{\circ}$, $\Theta=90^{\circ}$, $\Phi=0^{\circ}$. The early stages of the waveform are performed with large time steps to be able to model the full $600$ million year orbit. Each of the two subplots A and B are $\approx 7$ days of data. During this time B had $\approx 115$ peaks (near continuous emission) whereas A had 1. It is important to note that A, representing a gravitational wave ``peep" has an orbital period of $T\approx185000$ years at this early stage of the inspiral. The y-axis is the plus polarized strain amplitude $h_+$ scaled by D, the radial distance of the observation point from the source, and $\mu$ is the test body’s mass.}
  \label{inspiral}
\end{figure*}

Figure \ref{fig:waveform} shows a portion of the inspiral with an extremely large time period ($T\approx 185000$ years). As the orbit evolves through multiple cycles the period drastically decreases until the EMRI is near plunge where there is near continuous emission. An example of this can be seen in Figure \ref{inspiral} which shows a complete inspiral from the point of capture through plunge. The entirety of this inspiral lasts $t\approx6\times10^8$ years which is far longer than LISA's proposed observing time. This figure is included as a proof of concept to demonstrate the portion of the inspiral this paper is focused on. In order to fully model the very long inspiral, a larger time step was utilized for the earlier stages of the orbit. There are two subfigures included in Figure \ref{inspiral} which each show 7 days' worth of data. It is very clear to see the difference, as in (a) the orbital period is $T\approx185000$ years and is zoomed in to a portion around the gravitational wave peep. In subfigure (b), in the same amount of time there are roughly 115 peaks indicating near continuous gravitational wave emission. Subfigure (b) is very likely to be detectable as it will have a much higher signal-to-noise ratio. This project however is far more focused on subfigure (a), the gravitational wave peep. Since most EMRIs will remain in this long-period portion of their inspiral for a vast majority of their ``lifetime" it is possible that these small peeps may combine together in various ways and possibly obscure otherwise detectable sources such as those in subfigure (b). One feature of Figure \ref{inspiral} is especially worthy of note. The characteristic chirp profile of the inspiral is largely confined to the late stage of the waveform, at least in amplitude terms. The amplitude of the waveform increases sharply at the end of the inspiral but is roughly constant for most of the time elapsed. The reason is that a highly eccentric orbit evolved under radiation reaction chiefly by decreasing its apoapsis radius. The periapsis radius changes very little. It is this radius that determines the frequency and amplitude of the peep, which thus evolves very little over the course of the inspiral waveform. This will undoubtedly help in modeling the peep signal confusion noise.\par

\subsection{Window Functions}

\begin{figure}[h!]
  \begin{minipage}[b]{0.5\linewidth}
    \centering
    \includegraphics[width=\linewidth]{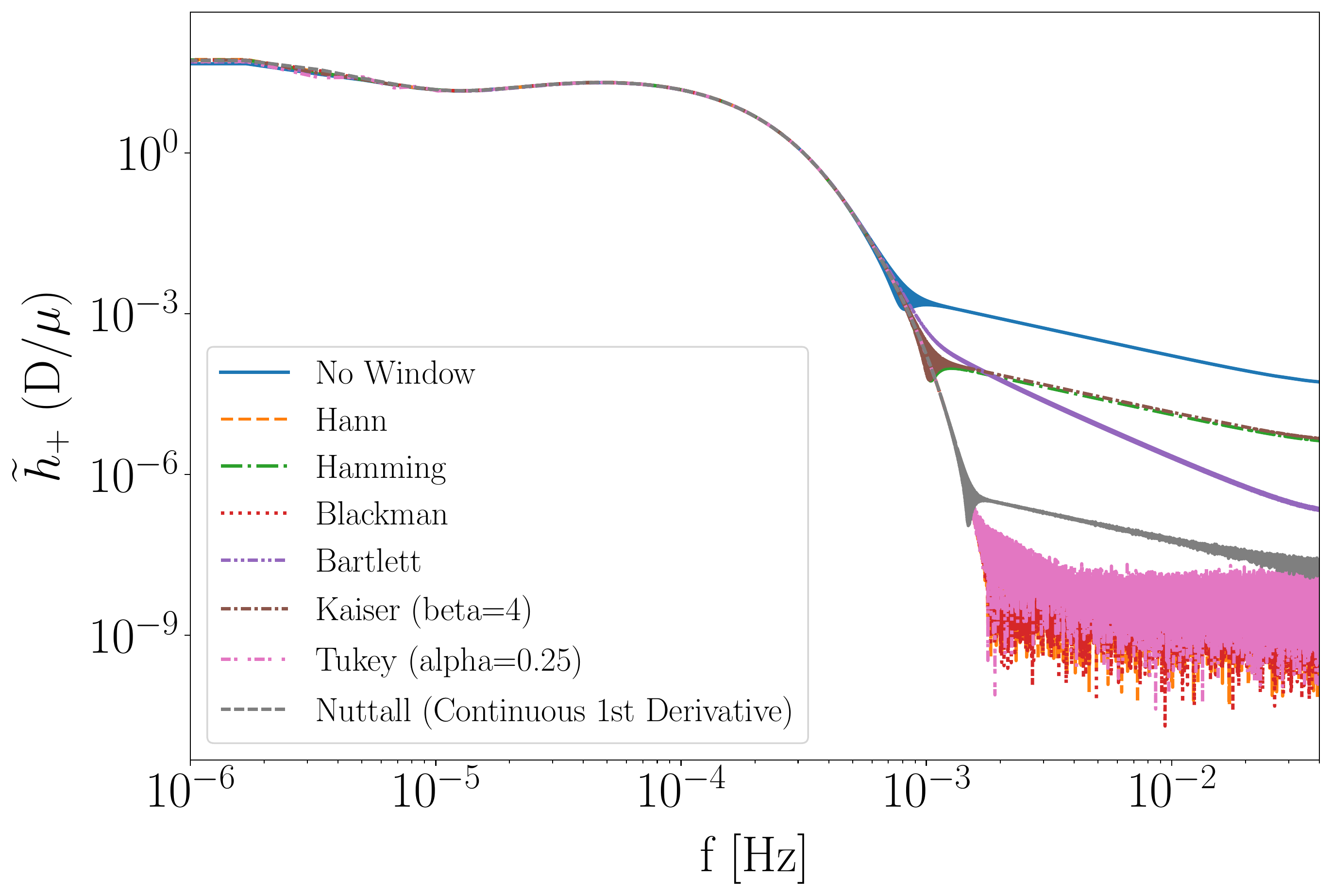}
  \end{minipage}%
  \begin{minipage}[b]{0.5\linewidth}
    \centering
    \includegraphics[width=\linewidth]{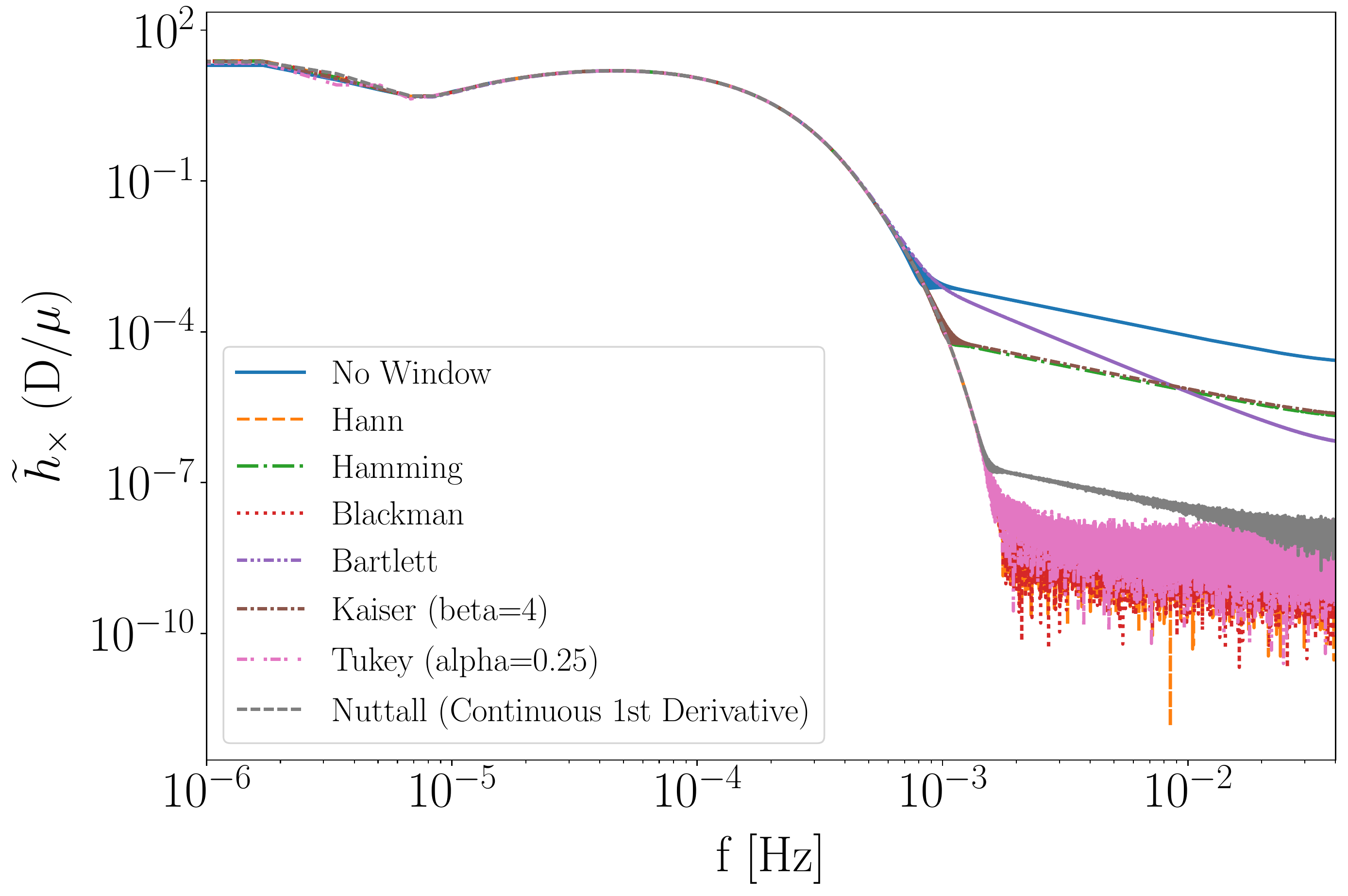}
  \end{minipage}
  \caption{FFT Spectra for $h_+$ (left) and $h_\times$ (right) with parameters $a=0.9M$, $p=120M$, $e=0.999999$, $\iota=0.0^{\circ}$, $dt=10$ seconds, $\Theta=0.0^{\circ}$, and $\Phi=90^{\circ}$ with several window functions outlined in section III-B overlaid on top of each other. The spectra for each window function overlap quite well for both the $h_+$ and $h_\times$ and it is only at the higher frequencies where there is any deviation. The Hann, Blackman, Tukey, and Nuttall windows seem to be the most effective at reducing the amplitude at these higher frequencies and thus these will be the windows utilized throughout this study.}
  \label{fig:Window}
\end{figure}

Window functions play an important role in signal processing, particularly in the context of spectral analysis. In our case where we are Fast Fourier Transforming a finite time-domain signal to the frequency-domain, they are often employed to reduce spectral leakage and control sidelobe levels. To implement the window function, we multiplied the time-domain signal by the window function. Several examples of window functions are utilized and shown on the same set of data for $h_+$ and $h_\times$ in Figure \ref{fig:Window}. There are several window functions available, each having there own unique characteristics. The rectangular window (no specific window function) is the simplest, but suffers from significant spectral leakage and high sidelobes, making it unsuitable. In contrast, the Hann \cite{blackman-tukey}, Hamming \cite{blackman-tukey}, and Blackman \cite{blackman-tukey} windows effectively suppress spectral leakage, lowering sidelobe levels and improving frequency resolution. The Hann window provides a trade-off between spectral leakage and main lobe width, whereas the Hamming window offers better sidelobe attenuation at the cost of wider main lobes. The Blackman window further enhances sidelobe suppression, but it has a wider main lobe than the other two. The Bartlett \cite{bartlett} window offers excellent main lobe resolution but with significant sidelobes. The Kaiser \cite{kaiser} window uses a beta parameter which allows an adjustment between main lobe width and sidelobe levels. The Tukey \cite{blackman-tukey} window, which is a variant of the Hann window, allows for a smooth tapering of the signal's edges, making it suitable for finite signals with abrupt transitions. The Nutall four-term window with continuous first derivative (Nuttall 1981 \cite{nuttall1981}) was included as an option due to being implemented in Berry \& Gair (2013) \cite{berryObservingGalaxyMassive2013a}. The Nuttall window is designed to address the issue of discontinuities in signal data and we found that it works particularly well in cases where several orbits appear such as Figure \ref{fig:Parabolic}. From Figure \ref{fig:Window} for these specific parameters the only change is in the higher frequencies and it appears as though the Hann, Blackman, Tukey, and Nuttall windows seem to be the most effective at reducing the amplitude of the sidelobes minimizing the spectral leakage.

\subsection{Peep vs Burst Spectra}

\begin{figure}[h!]
  \begin{minipage}[b]{0.5\linewidth}
    \centering
    \includegraphics[width=\linewidth]{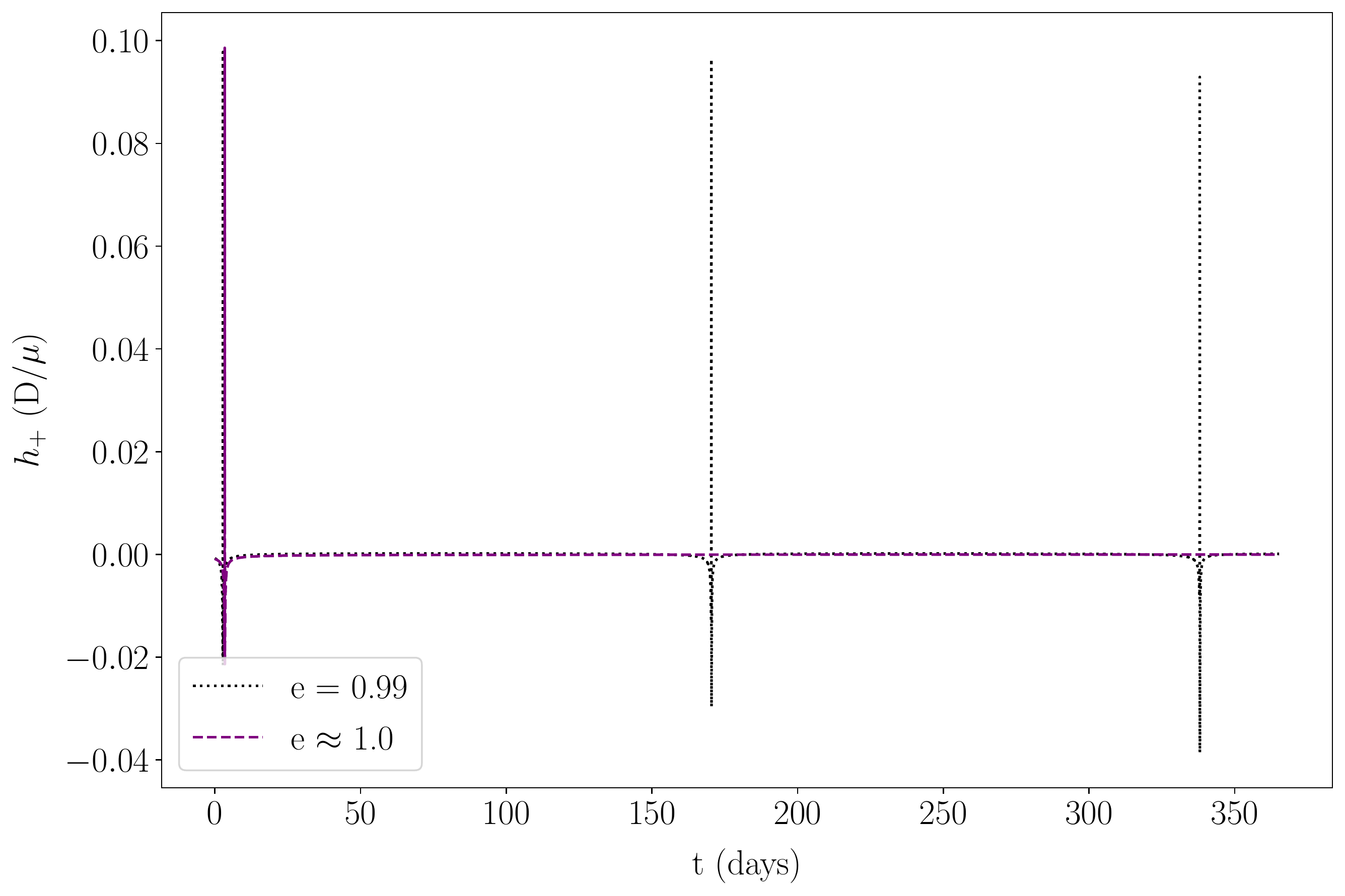}
  \end{minipage}%
  \begin{minipage}[b]{0.5\linewidth}
    \centering
    \includegraphics[width=\linewidth]{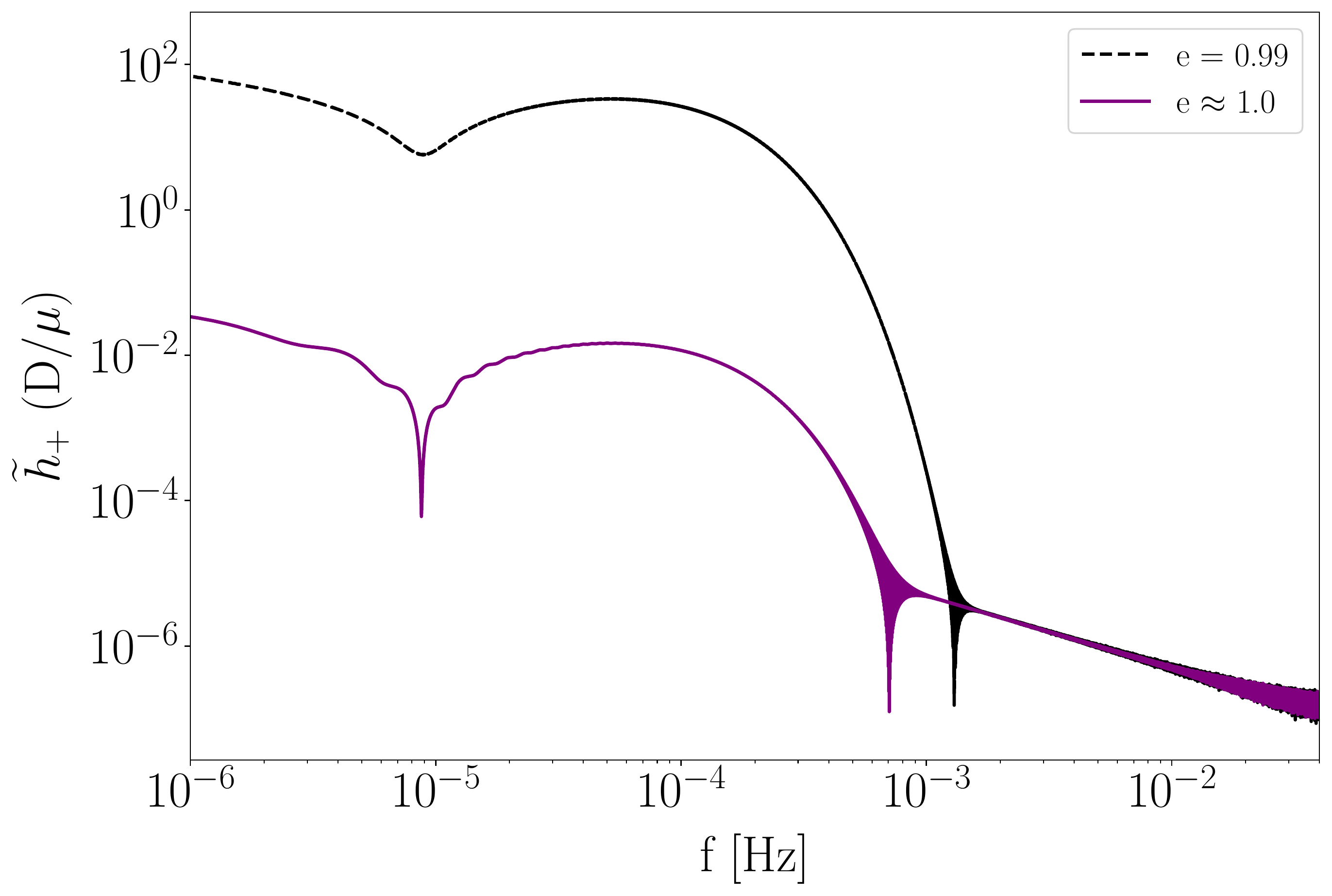}
  \end{minipage}
  \caption{(left): $h_+$ NK waveform with parameters $q=10^{-6}$, $a=0.9M$, $p=120M$, $\iota=0.0^{\circ}$, $dt=10$ seconds, $\Theta=0.0^{\circ}$, and $\Phi=90^{\circ}$ and eccentricities of $e=0.99$ (black dotted line) and $e=0.999999999\approx1$ (purple solid line) to show the difference between the lowest eccentricity in our study and one which is arbitrarily close to the parabolic case. Each case was run for a total of one year and the parabolic case emits gravitational radiation in a single burst at the beginning, while the $e=0.99$ case completes three orbits. This demonstrates the largest difference between the two regimes of the burst and the peep. This difference is even more prevalent when $p$ is much smaller, which in this same time period of 1 year with $p=15M$ the peep case exhibits $\approx50$ complete orbits compared to the parabolic case. (right): The $h_+$ FFT Spectra of the two waveforms using a Nuttall four-term window with continuous first derivative. The two biggest discrepancies are the amplitude and the frequency range. The $e=0.99$ case has an amplitude that peaks nearly 4 times larger than the parabolic case. In addition, the  $e=0.99$ case extends to higher frequencies than the parabolic case. The $h_\times$ case looks very similar to the $h_+$ and shows the same deviations between the parabolic burst case and the highly eccentric peep case.}
  \label{fig:Parabolic}
\end{figure}

Previous studies of EMRI/EMRB signal confusion noise have looked at the likely detectable signals with low eccentricity at the end of their ``lives" as well as the mostly undetectable bursts that only emit gravitational radiation once during LISA's lifespan. This study is focused on the long slow inspiral of every few centuries with an emission in the LISA bandwidth down to signals which emit gravitational radiation a few times per year. This is important to consider compared to the parabolic orbits of the EMRB signals as is demonstrated in Figure \ref{fig:Parabolic}. In both figures, we show the $h_+$ polarization for two signals, an EMRI with parameters $q=10^{-6}$, $a=0.9M$, $p=120M$, $e=0.99$, $\iota=0.0^{\circ}$, $\Theta=0.0^{\circ}$, and $\Phi=90^{\circ}$ in a black dotted line and the same parameters but with an EMRB parabolic orbit of $e=1.0$ shown in purple solid line. During a period of one year, the peep with $e=0.99$ crosses periapsis emitting gravitational radiation 3 times compared to the parabolic case which will only ever have one burst of gravitational radiation. In the right figure, we show the FFT spectra of the waveforms using a Nuttall four-term window with continuous first derivative and there are two very noticeable differences between the two spectra. The amplitude of the gravitational wave signal for the peeps in the FFT is $\approx4$ times larger than the amplitude of the parabolic burst. In addition to this, the peep extends to higher frequencies than the burst. This difference is even more extreme when looking at a closer approach. As an example, a system with $p=15M$ in the same period of 1 year exhibits $\approx50$ complete orbits. This causes an even larger difference in amplitudes in the frequency domain.\par

\subsection{Peep Spectra}

To determine the frequency of our EMRI waveforms, we take a Fast Fourier Transform (FFT) (using a Tukey window $\alpha=0.25$) of the gravitational wave peep and plot the values in the LISA frequency band ($0.1$ mHz - $0.1$ Hz). Figure \ref{fig:waveform}  shows the peep waveform for $h_+$ and $h_\times$ with initial parameters $q=10^{-6}$, $a=0.9M$, $p=120M$, $e=0.999999$, $\iota=25.0^{\circ}$, $\Theta=0.0^{\circ}$, $\Phi=90^{\circ}$, $dt=10s$ as well as the corresponding FFT spectra.\par

\begin{figure}[h!]
  \begin{minipage}[b]{0.5\linewidth}
    \centering
    \includegraphics[width=\linewidth]{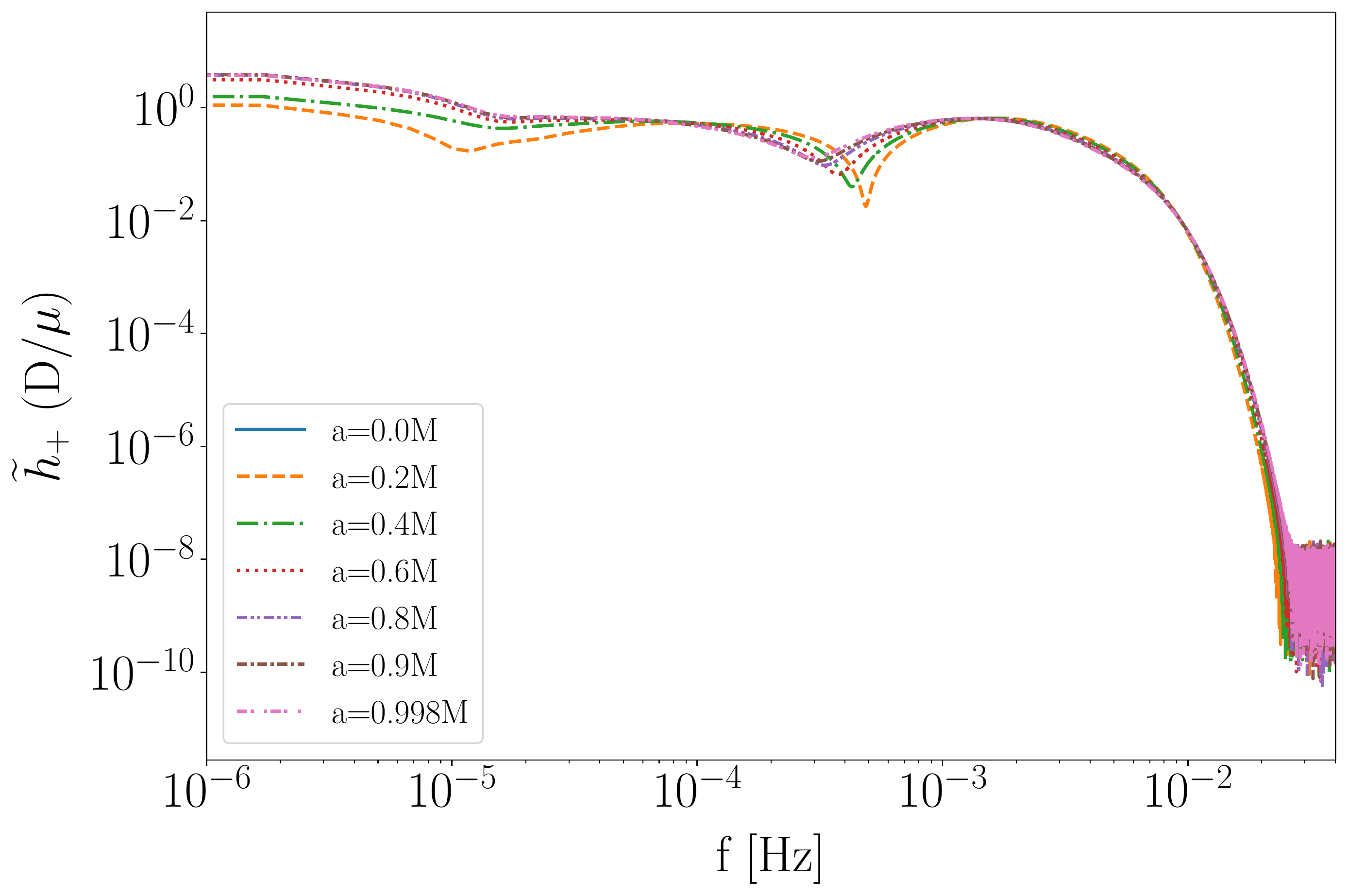}
  \end{minipage}%
  \begin{minipage}[b]{0.5\linewidth}
    \centering
    \includegraphics[width=\linewidth]{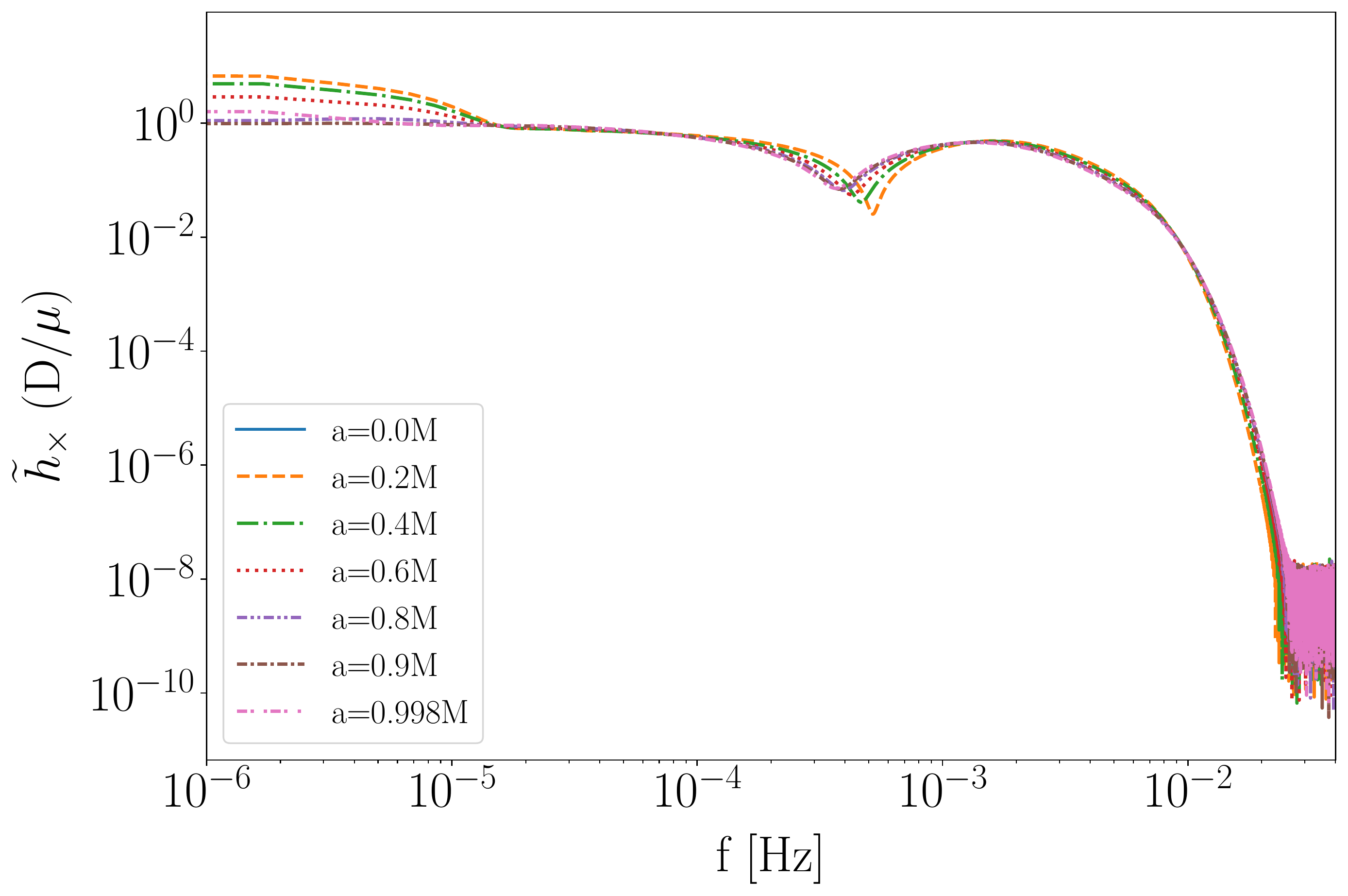}
  \end{minipage}
  \caption{FFT Spectra for $h_+$ (left) and $h_\times$ (right) using a Tukey Window with parameters $a=0.0-0.998M$, $p=15M$, $e=0.999999$, $\iota=0.0^{\circ}$, $dt=10$ seconds, $\Theta=0.0^{\circ}$, and $\Phi=90^{\circ}$ overlaid on top of each other. The spin values were chosen to cover the full range of no spin $a=0$ to a maximally spinning black hole $a=0.998$ and the semi-latus rectum value was chosen as our minimum value shown in Table \ref{tab:KParam} as that is when the spin of the MBH will have the greatest effect \cite{glampedakis_zoom_2002}. There is a slight shift in the lower frequencies where they are out of phase in $h_\times$, as well as the demarcation around $10^{-3}$ Hz which occurs in both $h_+$ and $h_\times$.}
  \label{fig:SpecCompA}
\end{figure}

\begin{figure*}
  \begin{subfigure}[b]{0.5\linewidth}
    \centering
    \includegraphics[width=0.975\linewidth]{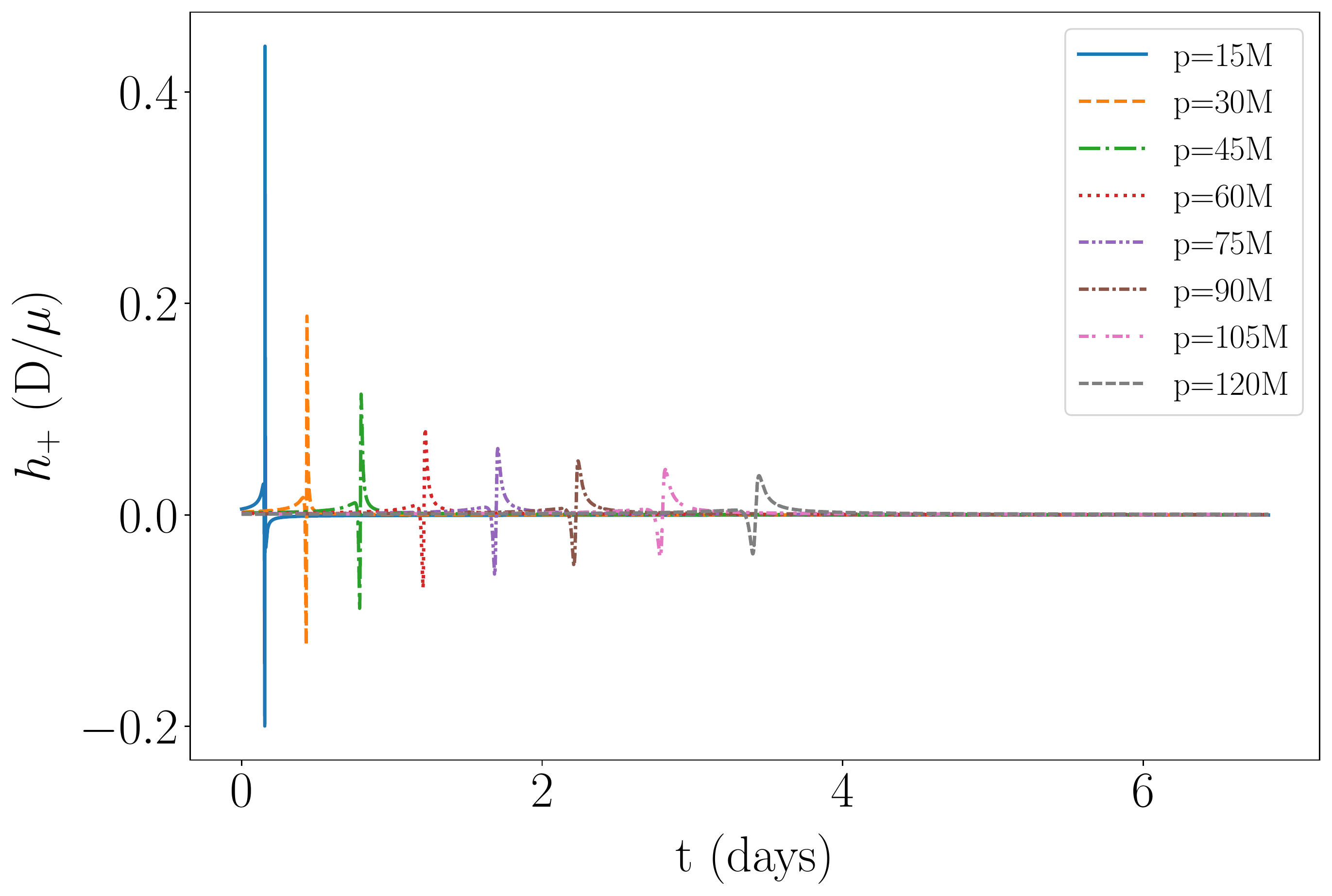} 
    \caption{$h_+$ Waveform}
  \end{subfigure}
  \begin{subfigure}[b]{0.5\linewidth}
    \centering
    \includegraphics[width=\linewidth]{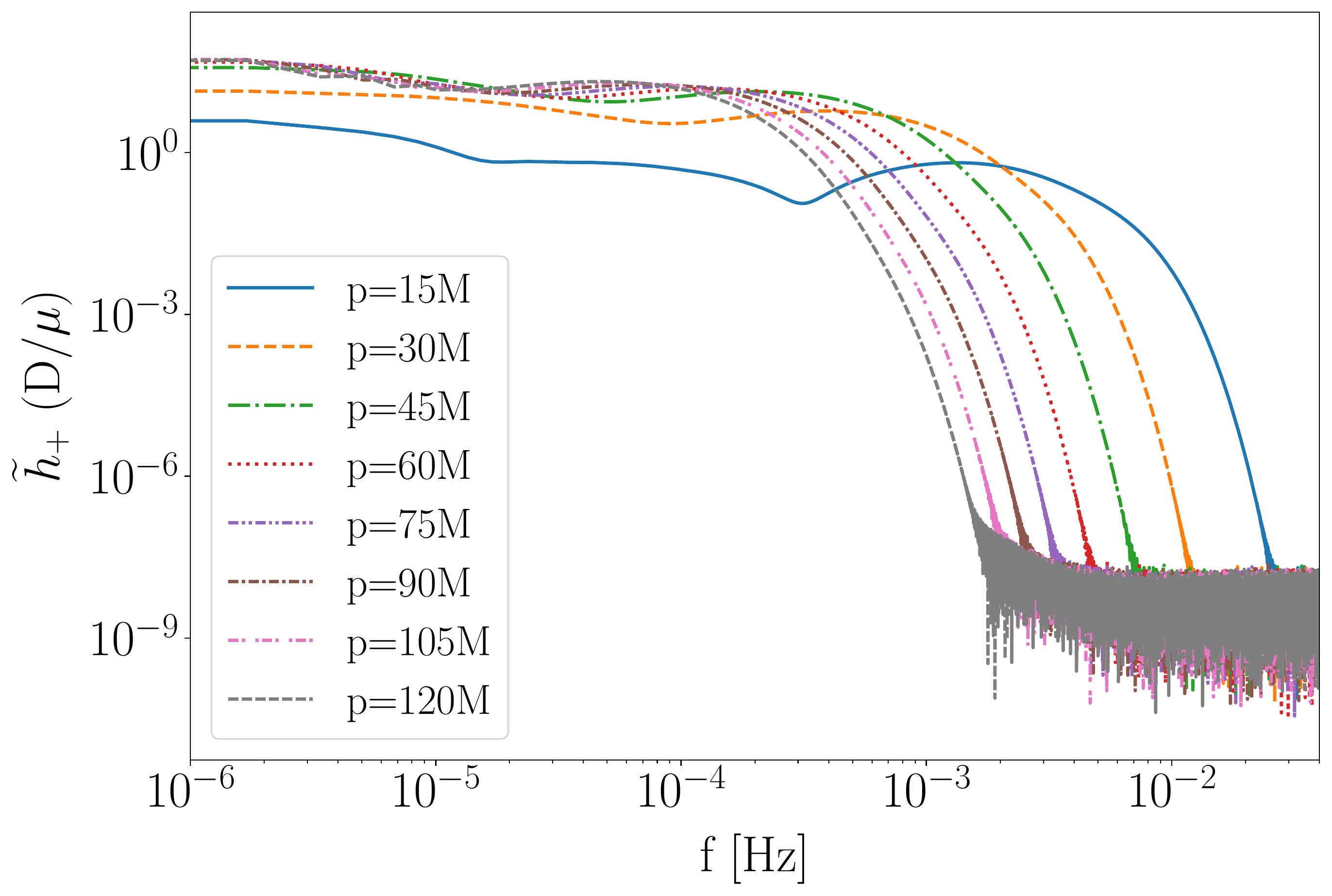}6
    \caption{$h_+$ Spectra}
  \end{subfigure} 
  \begin{subfigure}[b]{0.5\linewidth}
    \centering
    \includegraphics[width=0.975\linewidth]{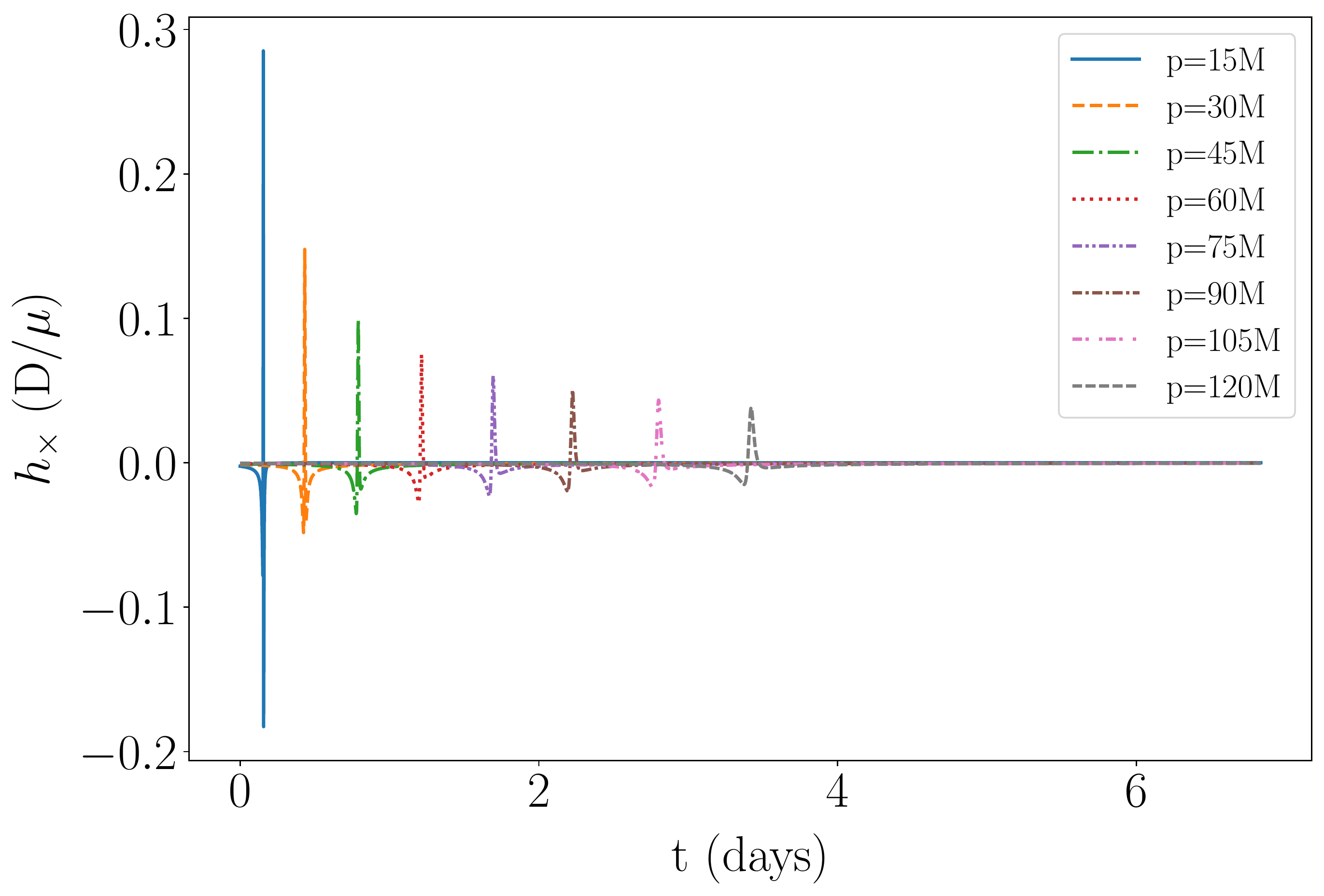} 
    \caption{$h_\times$ Waveform}
  \end{subfigure}
  \begin{subfigure}[b]{0.5\linewidth}
    \centering
    \includegraphics[width=\linewidth]{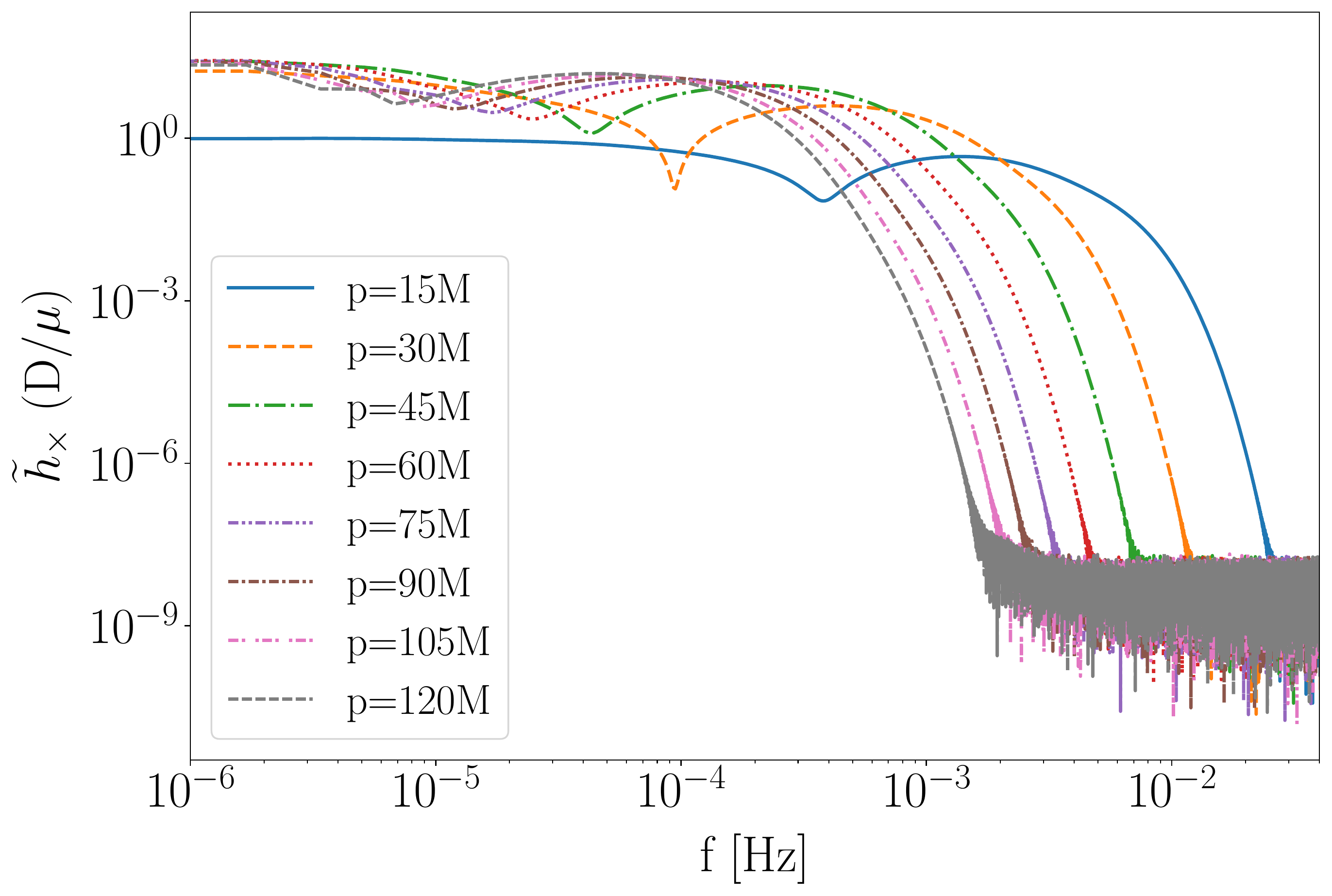}
    \caption{$h_\times$ Spectra}
  \end{subfigure}  
  \caption{Waveform and FFT Spectra for $h_+$ (top) and $h_\times$ (bottom) with parameters $a=0.9M$, $p=15M-120M$, $e=0.999999$, $\iota=0.0^{\circ}$, $dt=10$ seconds, $\Theta=0.0^{\circ}$, and $\Phi=90^{\circ}$ overlaid on top of each other. The semi-latus rectum values were chosen based on the capture parameters in Table \ref{tab:KParam}. In (a) and (c), showing the waveforms, it is evident that the amplitude of the peep is larger for smaller values of $p$. However, in (b) and (d), showing the spectra, lower $p$ values correspond to a smaller amplitude but higher peak frequencies. When looking at the extremes, it is notable that the peak amplitude of $p=120M$ is larger, while $p=15M$ exhibits contributions across a broader frequency range (Note: this is only showing a single peep in a 1-week sample compared to Figure \ref{fig:Parabolic} which was 1-year where there would be significantly more peeps in the $p=15M$ case, leading to a much larger overall amplitude). This broader frequency range implies a higher total energy, aligning with the expectations set by the waveforms in (a) and (c).}
  \label{fig:SpecCompP}
\end{figure*}

\begin{figure}[h!]
  \begin{minipage}[b]{0.5\linewidth}
    \centering
    \includegraphics[width=\linewidth]{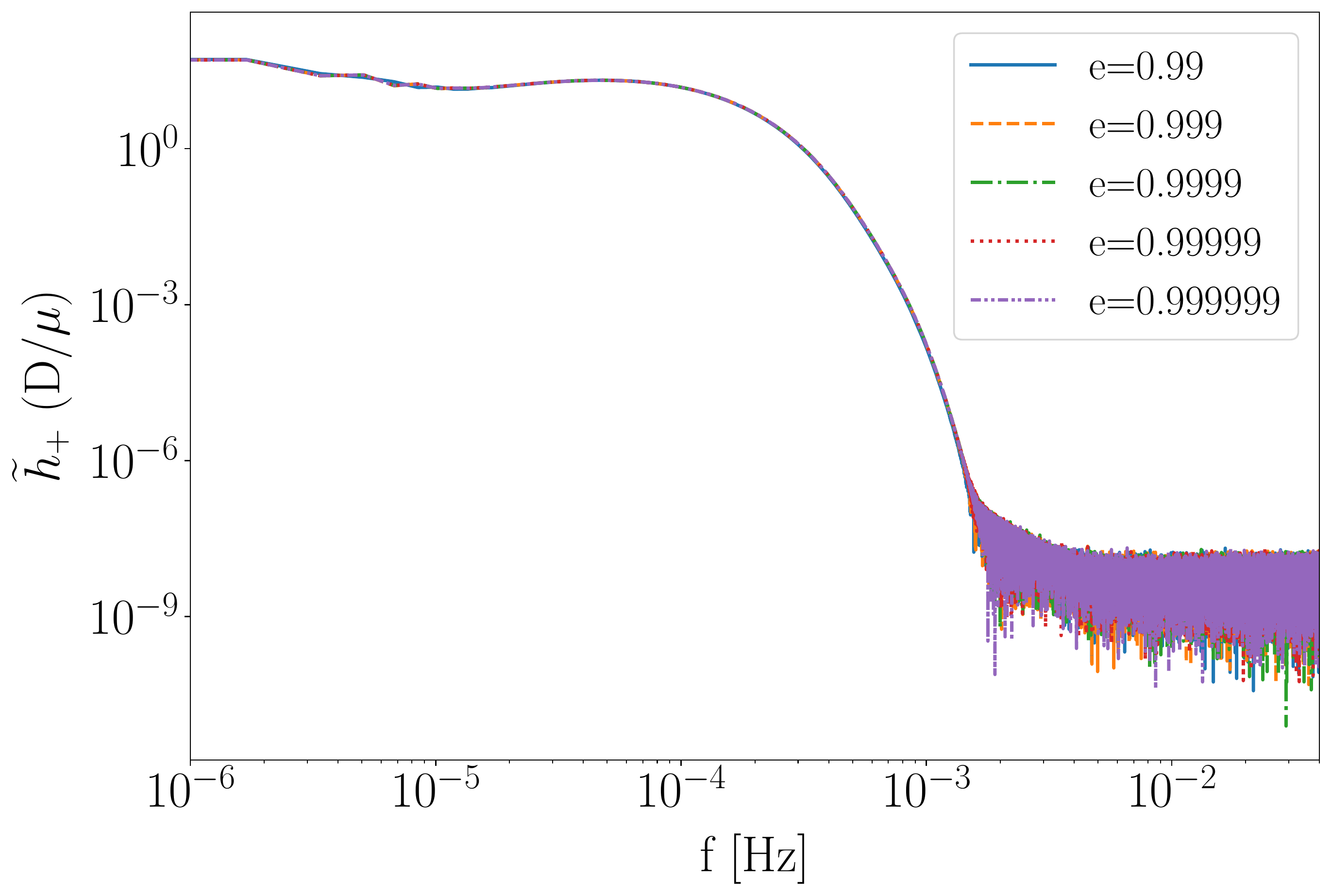}
  \end{minipage}%
  \begin{minipage}[b]{0.5\linewidth}
    \centering
    \includegraphics[width=\linewidth]{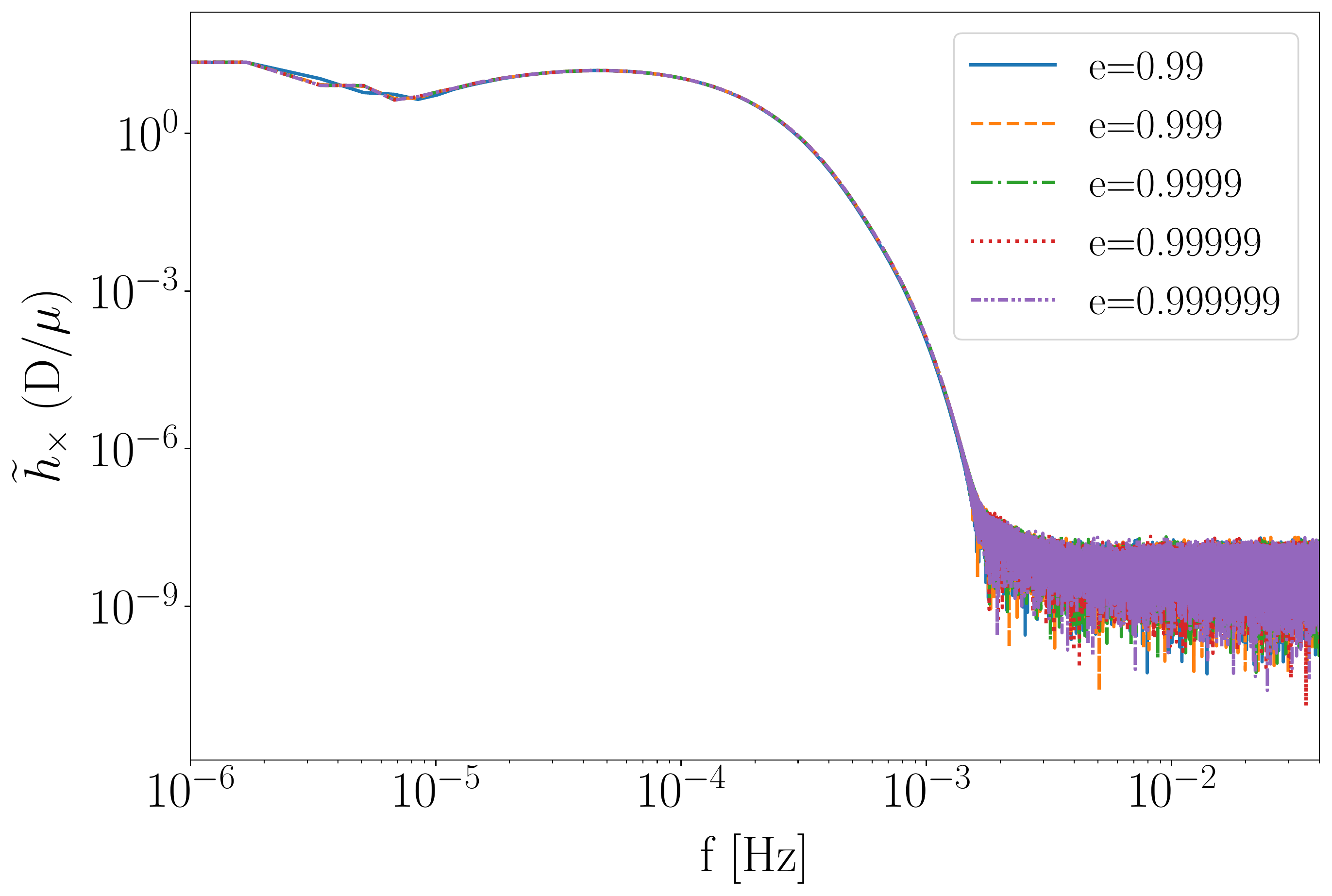}
  \end{minipage}
  \caption{FFT Spectra for $h_+$ (left) and $h_\times$ (right) using a Tukey Window with parameters $a=0.9M$, $p=120M$, $e=0.99-0.999999$, $\iota=0.0^{\circ}$, $dt=10$ seconds, $\Theta=0.0^{\circ}$, and $\Phi=90^{\circ}$ overlaid on top of each other. The eccentricity values were chosen based on the parameters in Table \ref{tab:KParam}. The eccentricity at this high of a range doesn't seem to significantly impact the resulting spectra leading to a consistent overlap and all values besides $e=0.99$ appear to match that of Figure \ref{fig:SpecCompA}.}
  \label{fig:SpecCompEcc}
\end{figure}

\begin{figure}[h!]
  \begin{minipage}[b]{0.5\linewidth}
    \centering
    \includegraphics[width=\linewidth]{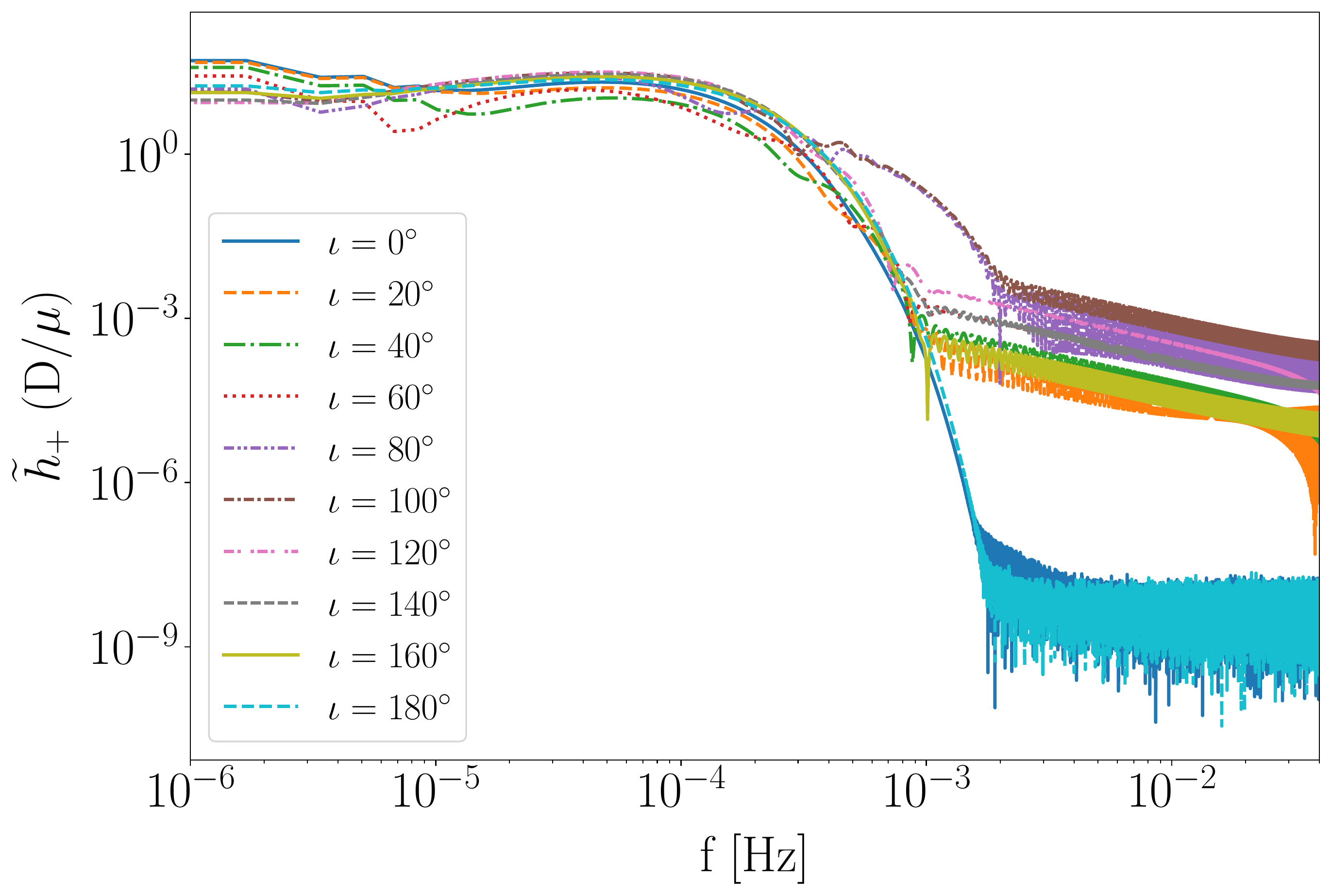}
  \end{minipage}%
  \begin{minipage}[b]{0.5\linewidth}
    \centering
    \includegraphics[width=\linewidth]{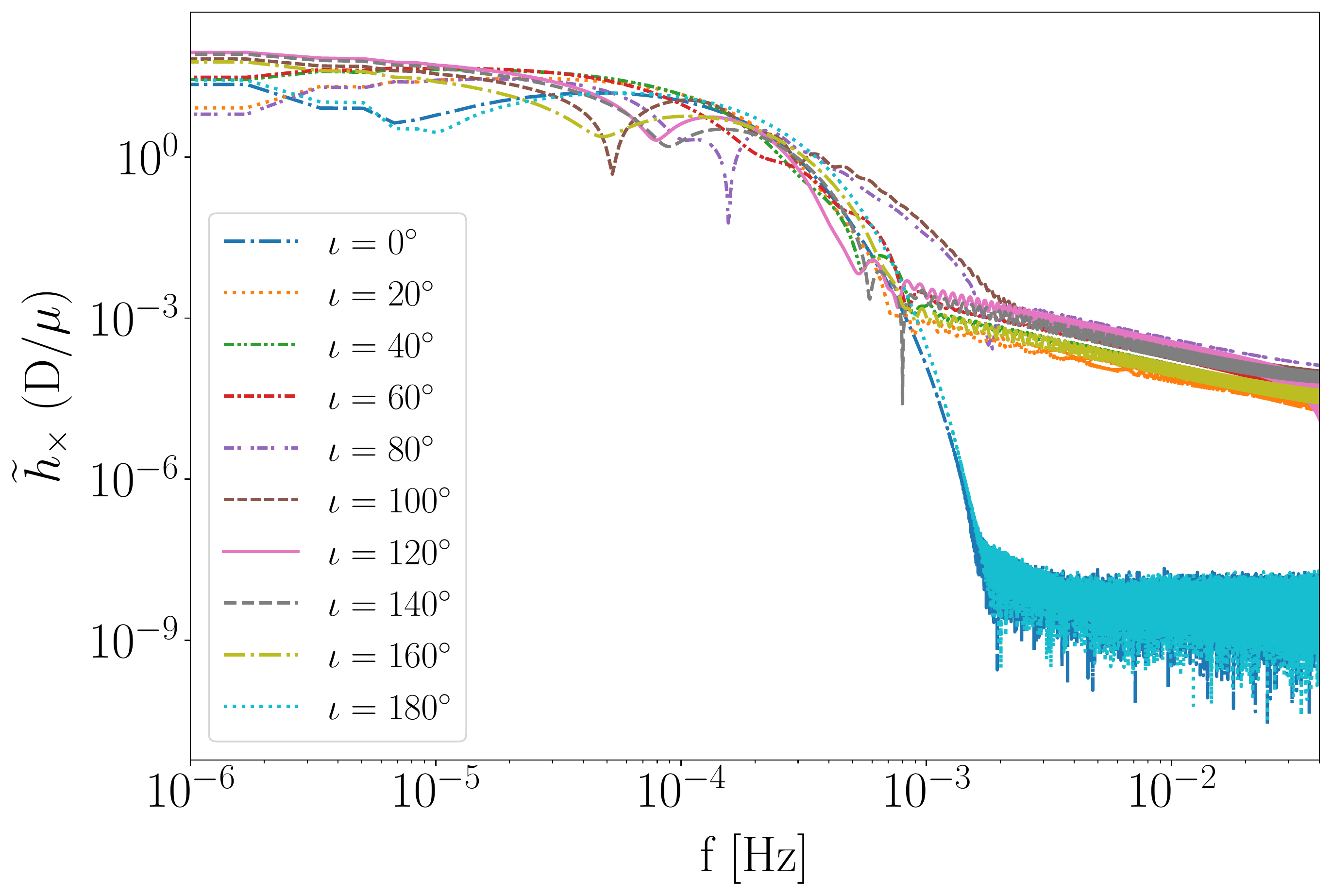}
  \end{minipage}
  \caption{FFT Spectra for $h_+$ (left) and $h_\times$ (right) using a Tukey Window with parameters $a=0.9M$, $p=120M$, $e=0.999999$, $\iota=0.0^{\circ}-180.0^{\circ}$, $dt=10$ seconds, $\Theta=0.0^{\circ}$, and $\Phi=90^{\circ}$ overlaid on top of each other. The inclination values were chosen to cover the full range of prograde ($0^\circ\leq\iota<90^\circ$) and retrograde ($90^\circ<\iota\leq180^\circ$) orbits. The results show the equatorial cases where $\iota=0^\circ,180^\circ$ are very similar to the previous figures. The interesting structure visible in the spectra of inclined orbits is worthy of further study. It is worth noting that the $\iota=80^\circ$ and $\iota=100^\circ$ could represent non-physical structures as they are very close to $\iota=90^\circ$ where the numerical kludge scheme breaks.}
  \label{fig:SpecCompIota}
\end{figure}

To better understand the relationship between the peep parameters and the corresponding spectra we have plotted the effects of changing those orbital parameters for $h_+$ and $h_\times$ in Figure \ref{fig:SpecCompA} (spin $a$), Figure \ref{fig:SpecCompP} (semi-latus rectum $p$), Figure \ref{fig:SpecCompEcc} (eccentricity $e$), and Figure \ref{fig:SpecCompIota} (inclination $\iota$). For the change in spin, it was found to be best to use the parameters $a=0.0-0.998M$, $p=15M$, $e=0.999999$, $\iota=0.0^{\circ}$, $\Theta=0.0^{\circ}$, and $\Phi=90^{\circ}$. With a small semi-latus rectum, the spin of the MBH will have a greater impact on the orbit on the EMRI, and thus the smallest $p$ from Table \ref{tab:KParam} was chosen, while also expanding the spin parameters to cover the full range from Schwarzschild ($a=0$) to maximally spinning ($a=0.998$). For the higher $p$ values the impact of the spin parameter was much less. One important difference between the $h_+$ and $h_\times$ is that in the $h_\times$ the peak becomes much more noticeably demarcated from the rest of the orbit compared to the $h_+$ plot.\par

For Figure \ref{fig:SpecCompP}, we adjusted the semi-latus rectum $15M<p<120M$ while maintaining fixed parameters of $a=0.9M$, $e=0.999999$, and $\iota=0.0^{\circ}$. The results of the spectra show that as p decreases, the peak frequency increases. The amplitudes remain fairly consistent for all $p$ values except for $p=15M$ where there is a decrease in amplitude throughout for both the $h_+$ and $h_\times$. For the $h_+$ the spectra have a noticeable dip before it increases into the peak frequency. For $h_\times$ this only occurs in the $p=15M$ and $p=30M$ cases. What this figure shows is how much the semi-latus rectum plays a role in the peak frequency.\par

Altering eccentricity, while keeping a consistent $a=0.9$, $p=120M$, and $\iota=0^{\circ}$ produces less noticeable changes and the spectra for various similar orbits of different eccentricities are barely distinguishable for $0.99<e<0.999999$ for both $h_+$ and $h_\times$ as shown in Figure \ref{fig:SpecCompEcc}. These parameters were chosen based on the estimated capture parameters detailed in Table \ref{tab:KParam}. For the eccentricity ranges we are looking at there is not likely to be a significant deviation between the low end of our eccentricity and the higher end.\par

The orbital inclination is adjusted in Figure \ref{fig:SpecCompIota} covering the range of $\iota=0.0^{\circ}-180.0^{\circ}$ which includes prograde ($0^\circ\leq\iota<90^\circ$) and retrograde ($90^\circ<\iota\leq180^\circ$) orbits while maintaining a fixed $a=0.9M$, $p=120M$, $e=0.999999$. The resulting spectra for the equatorial cases $\iota={0^\circ,180^\circ}$ follow the same trends as previous figures. The inclined orbits, however, have a much more interesting structure with multiple maxima and a larger amplitude at higher frequencies which is worthy of further study. Utilizing several windowing functions results in the inclined orbits maintaining their larger amplitudes which implies that it is a physical structure. It is important to point out that the maxima between $10^{-4}-10^{-2}$Hz in $\iota=80^\circ$ and $\iota=100^\circ$ could be non-physical as the numerical kludge scheme breaks at $\iota=90^\circ$. \par

\subsection{Adjusting the Observing angles $\Theta$ and $\Phi$}

\begin{figure*}
  {\large Adjusting Viewing Angle $\Phi$ for $a=0.9M$, $p=120M$, $e=0.999999$, $\iota=0^{\circ}$, and $\Theta=0^{\circ}$}\par\medskip
  \begin{subfigure}[b]{0.5\linewidth}
    \centering
    \includegraphics[width=\linewidth]{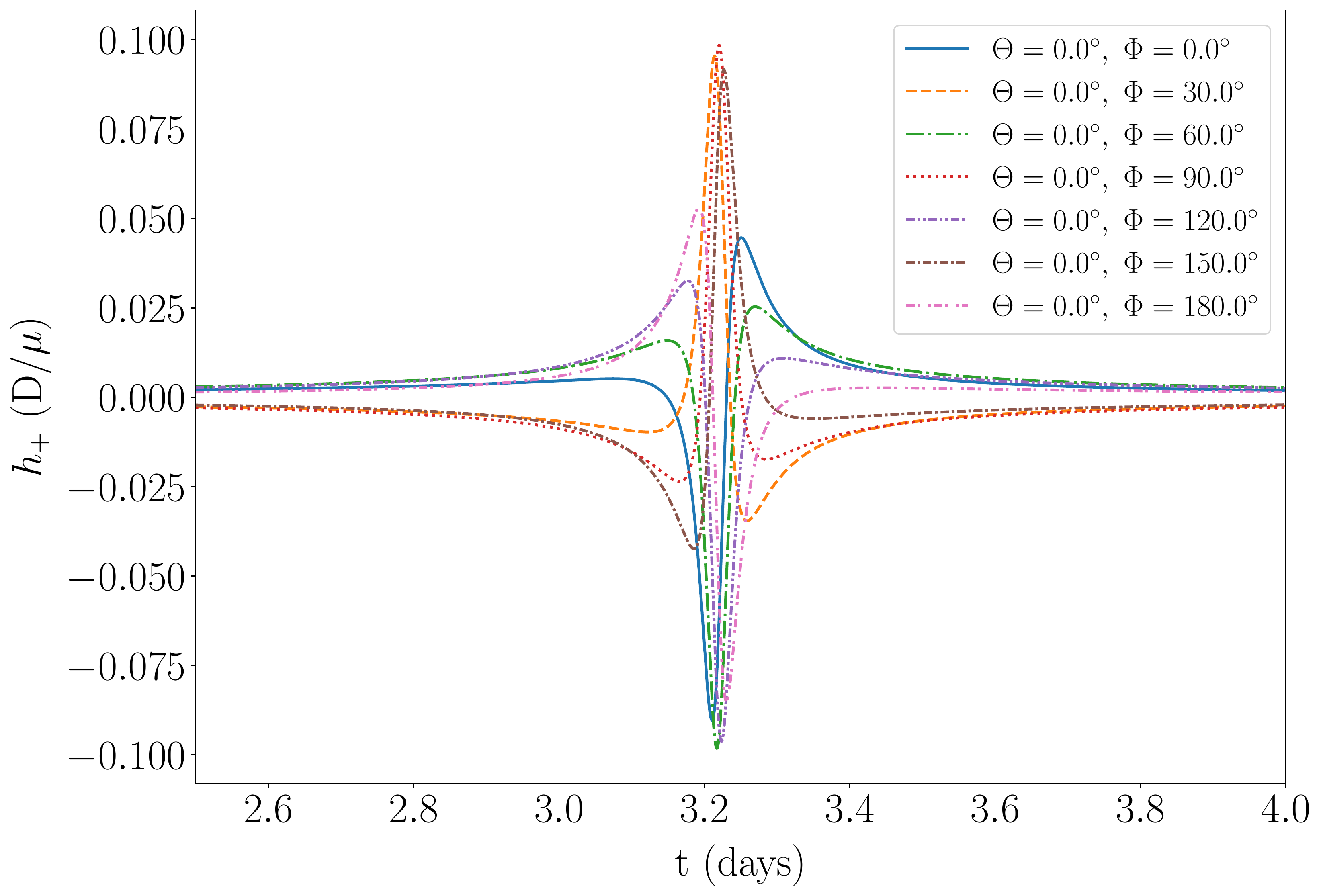} 
    \caption{$h_+$ Waveform}
  \end{subfigure}
  \begin{subfigure}[b]{0.5\linewidth}
    \centering
    \includegraphics[width=\linewidth]{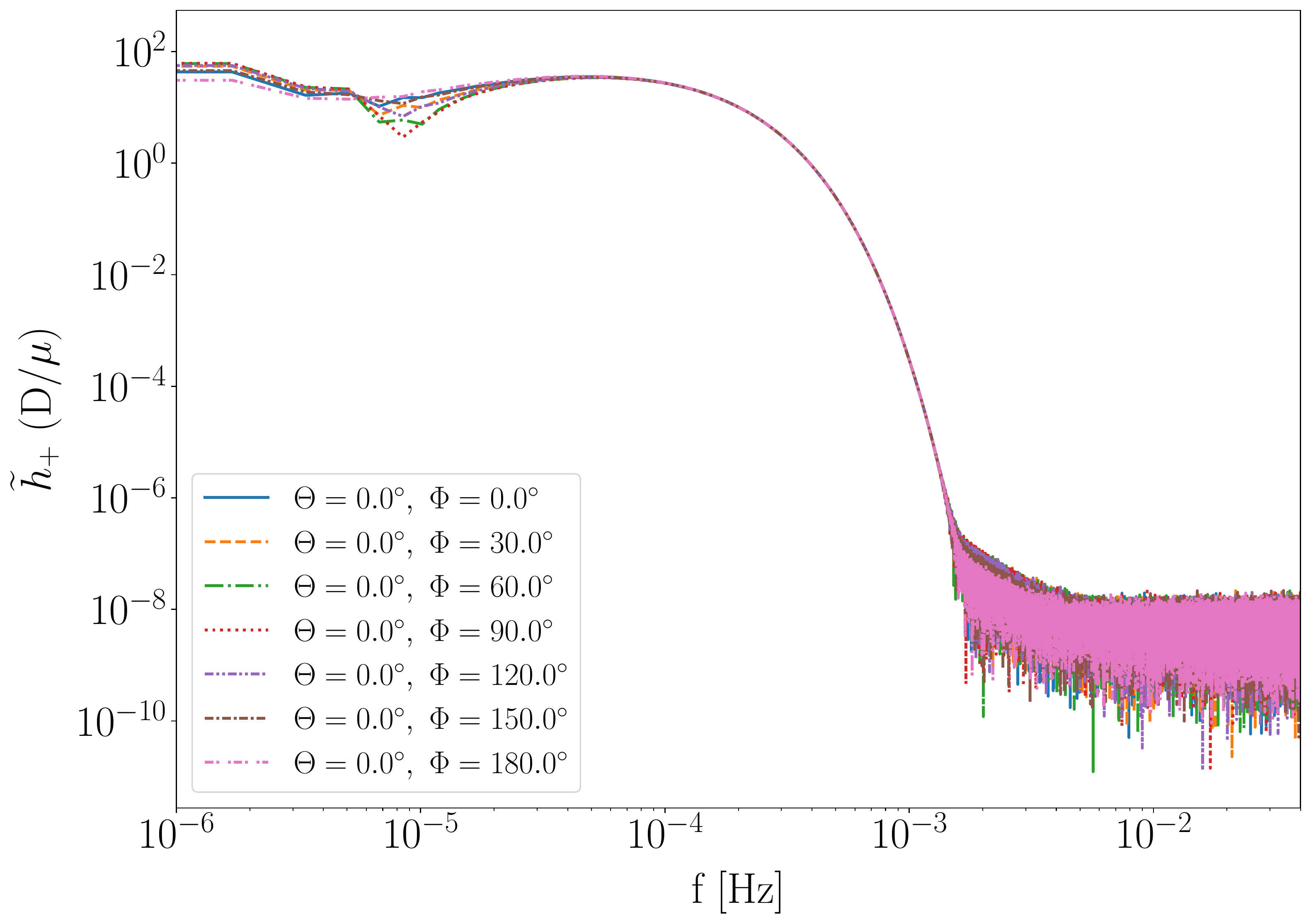}
    \caption{$h_+$  Spectra}
  \end{subfigure} 
  \begin{subfigure}[b]{0.5\linewidth}
    \centering
    \includegraphics[width=\linewidth]{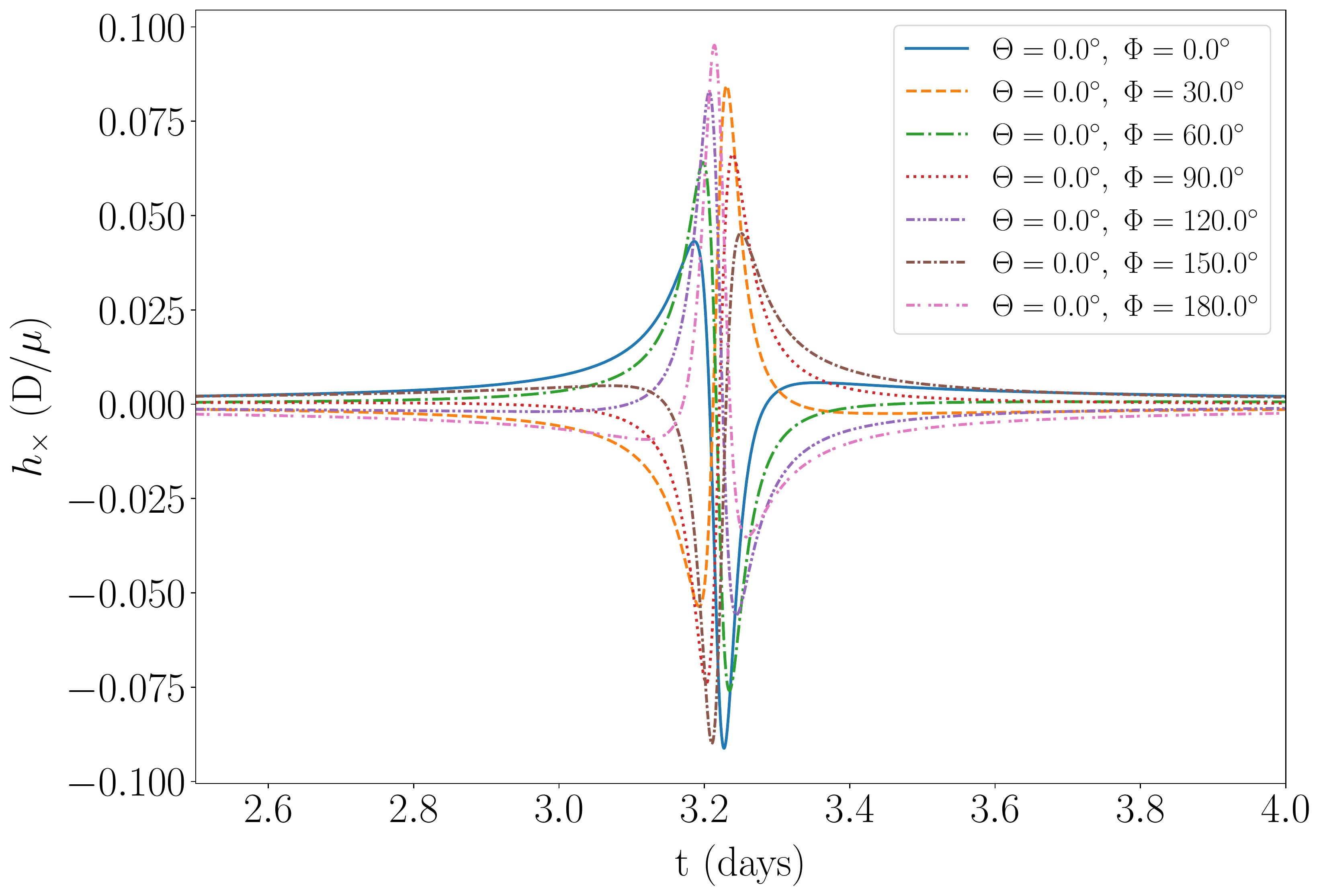} 
    \caption{$h_\times$ Waveform}
  \end{subfigure}
  \begin{subfigure}[b]{0.5\linewidth}
    \centering
    \includegraphics[width=\linewidth]{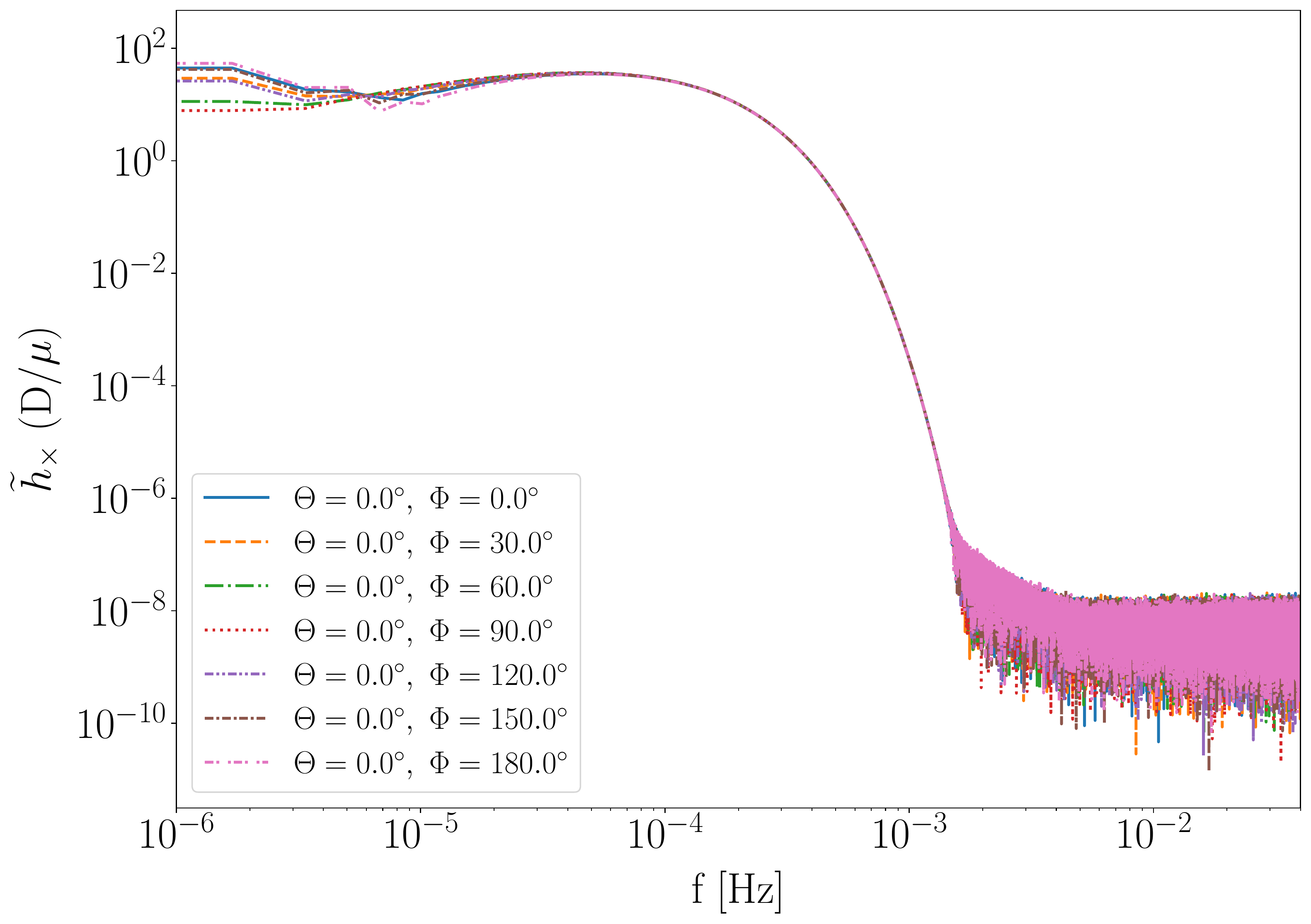}
    \caption{$h_\times$ Spectra}
  \end{subfigure} 
  \caption{Numerical Kludge waveform created with $a=0.9M$, $p=120M$, $e=0.999999$, $\iota=0.0^{\circ}$, $\Theta=0^{\circ}$ while adjusting the azimuthal viewing angle $\Phi$ from $0^\circ$ to $180^\circ$. In (a) and (c), we are showing the $h_+$ and $h_\times$ waveforms respectively while adjusting the azimuthal angle. This causes a slight phase shift for each of the waveforms. In (b) and (d) we show the corresponding FFT spectra using a Tukey Window with very little discernible change between the two indicating that adjusting the azimuthal viewing angle does not seem to affect the outgoing signal.}
   \label{theta0} 
\end{figure*}

\begin{figure*}
  {\large Adjusting Viewing Angle $\Phi$ for $a=0.9M$, $p=120M$, $e=0.999999$, $\iota=0^{\circ}$, and $\Theta=90^{\circ}$}\par\medskip
  \begin{subfigure}[b]{0.5\linewidth}
    \centering
    \includegraphics[width=\linewidth]{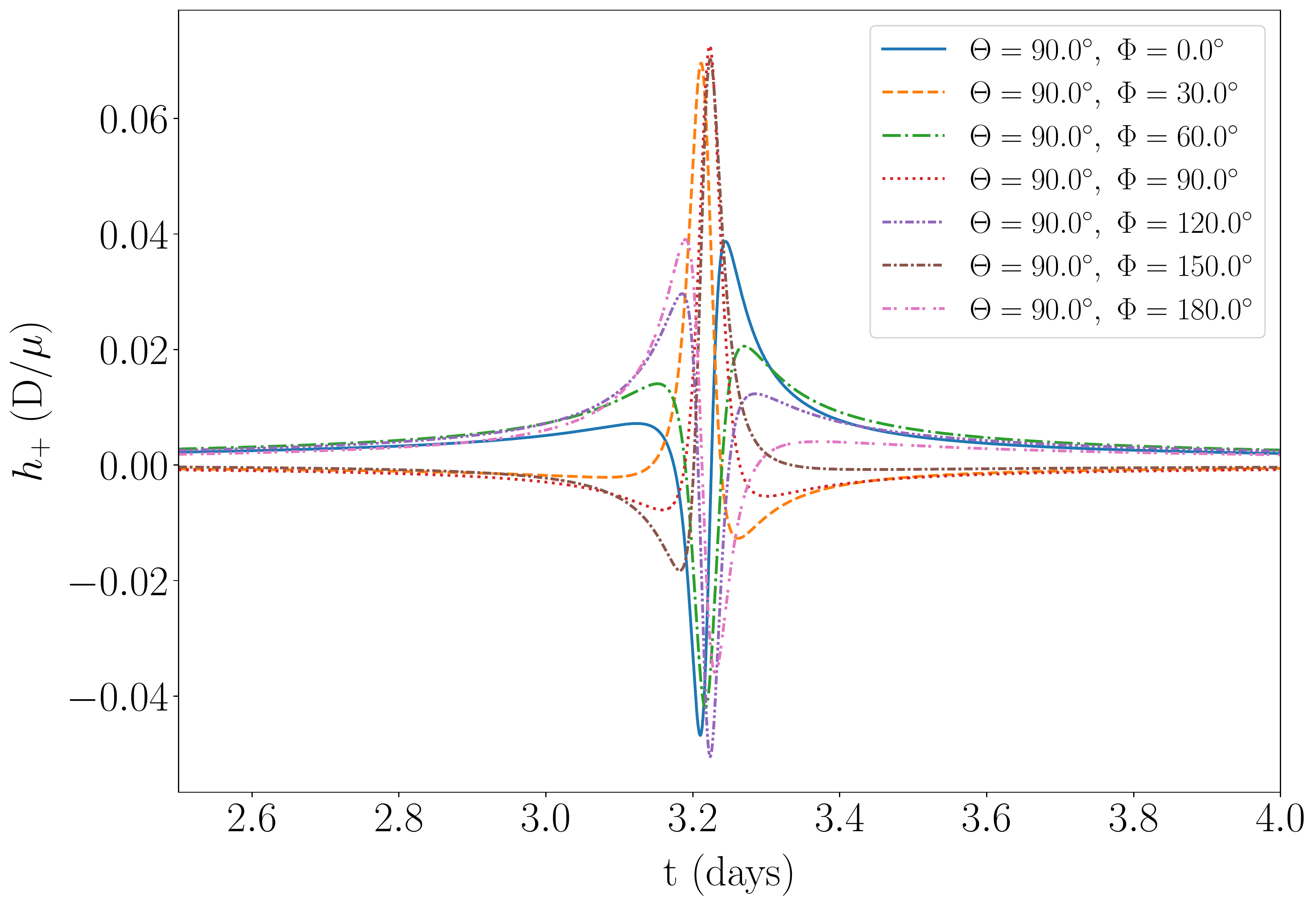} 
    \caption{$h_+$ Waveform}
  \end{subfigure}
  \begin{subfigure}[b]{0.5\linewidth}
    \centering
    \includegraphics[width=\linewidth]{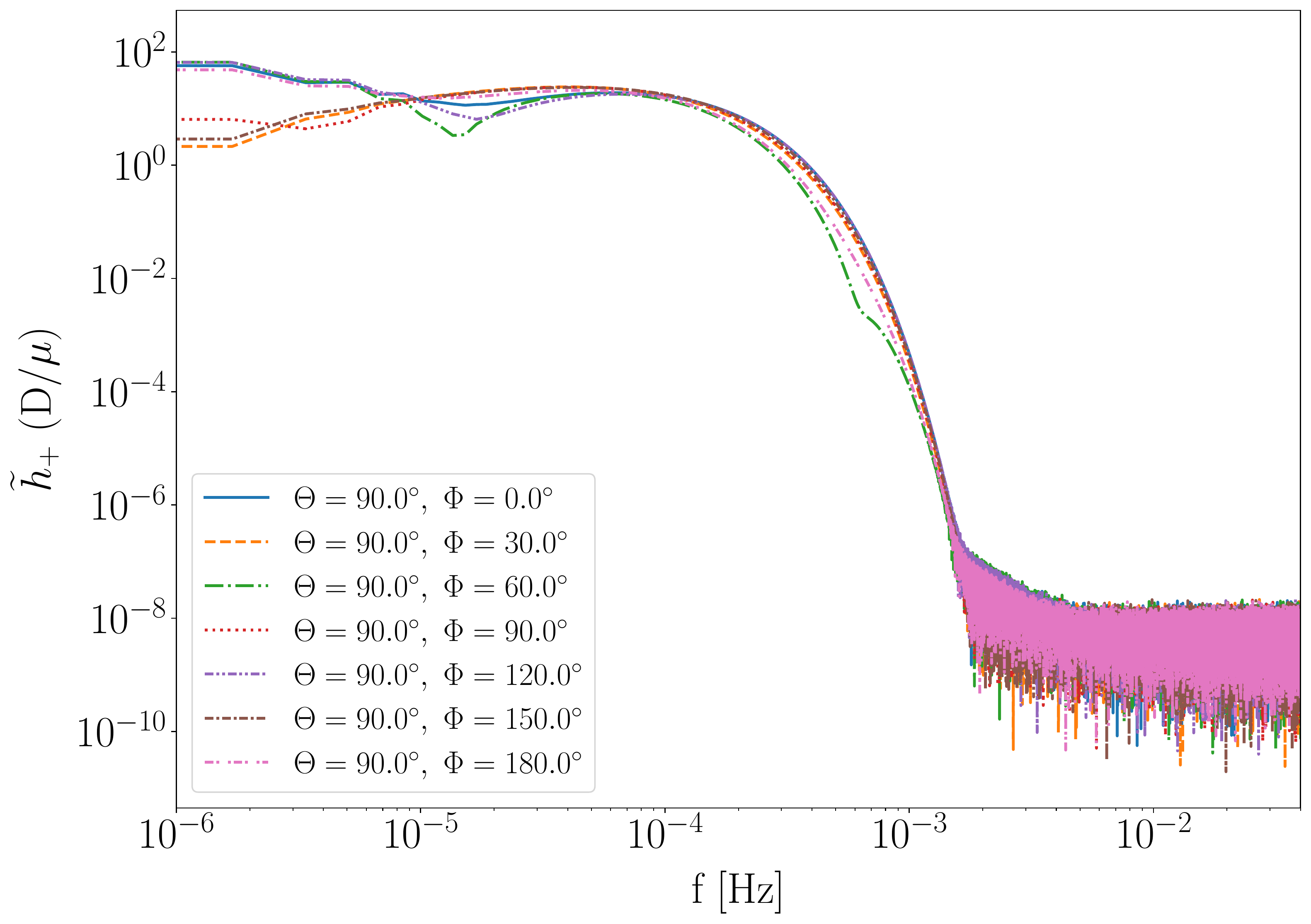}
    \caption{$h_+$ Spectra}
  \end{subfigure} 
  \begin{subfigure}[b]{0.5\linewidth}
    \centering
    \includegraphics[width=\linewidth]{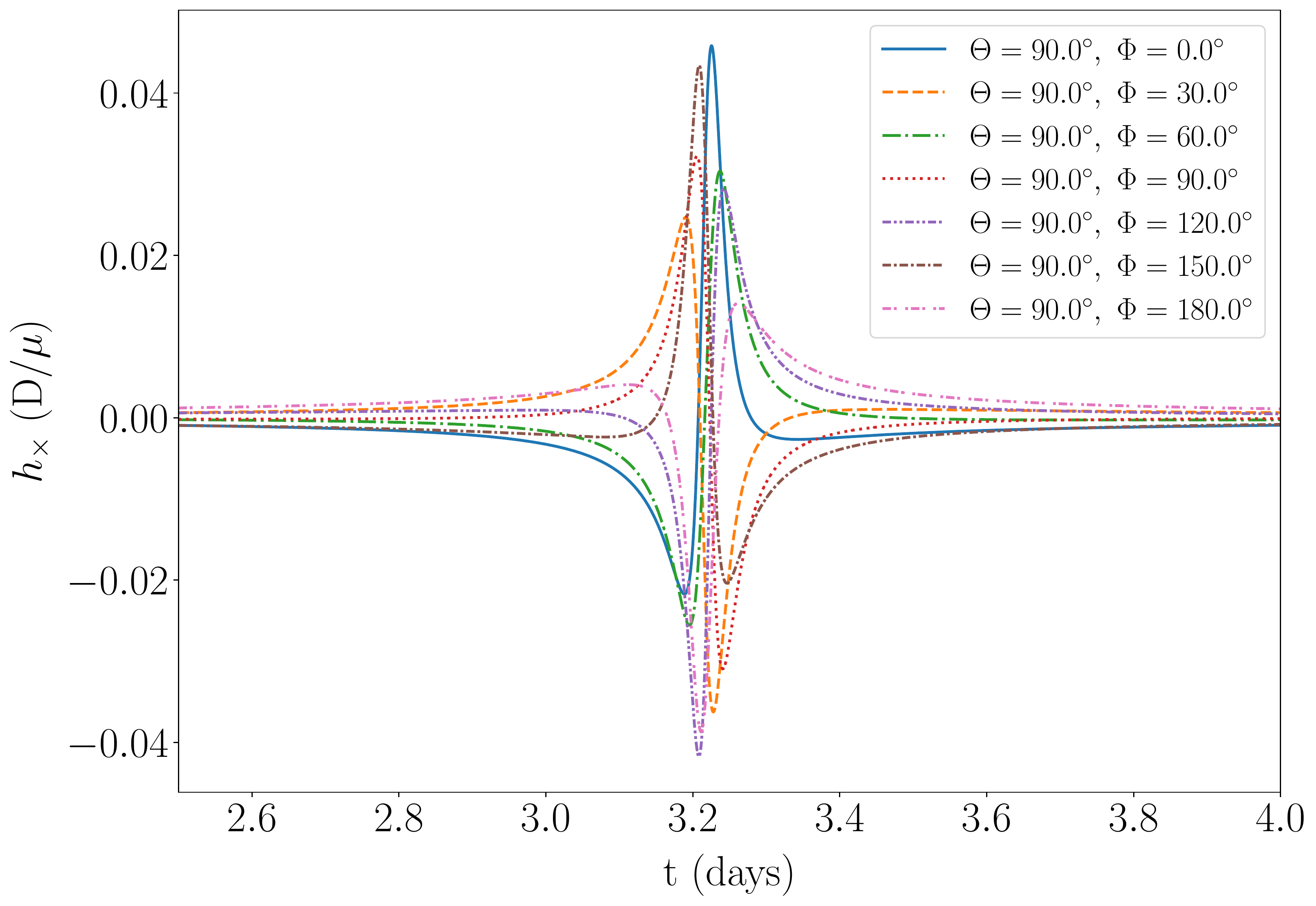} 
    \caption{$h_\times$ Waveform}
  \end{subfigure}
  \begin{subfigure}[b]{0.5\linewidth}
    \centering
    \includegraphics[width=\linewidth]{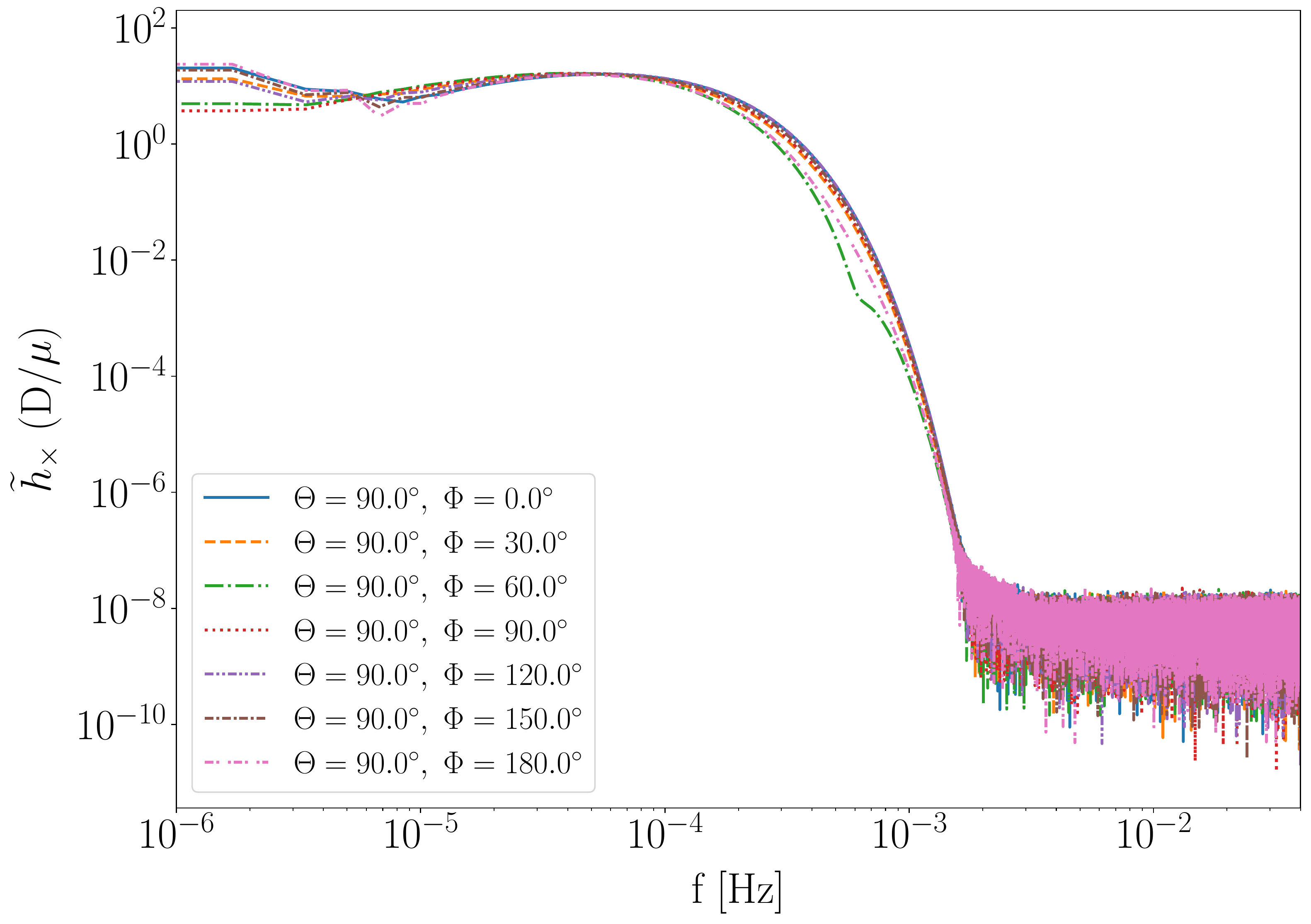}
    \caption{$h_\times$ Spectra}
  \end{subfigure} 
  \caption{Numerical Kludge waveform created with $a=0.9M$, $p=120M$, $e=0.999999$, $\iota=0.0^{\circ}$, $\Theta=90^{\circ}$ while adjusting the azimuthal viewing angle $\Phi$ from $0^\circ$ to $180^\circ$. In (a) and (c), we are showing the $h_+$ and $h_\times$ waveforms respectively. In (b) and (d) we show the corresponding FFT spectra using a Tukey Window and just as in Figure \ref{theta0}  there is very little deviation throughout the spectra, although the lower frequencies do show a much more discernable separation. }
   \label{theta90} 
\end{figure*}

\begin{figure*}
  {\large Adjusting Viewing Angle $\Theta$ for $a=0.9M$, $p=120M$, $e=0.999999$, $\iota=0^{\circ}$, and $\Phi=0^{\circ}$}\par\medskip
  \begin{subfigure}[b]{0.5\linewidth}
    \centering
    \includegraphics[width=\linewidth]{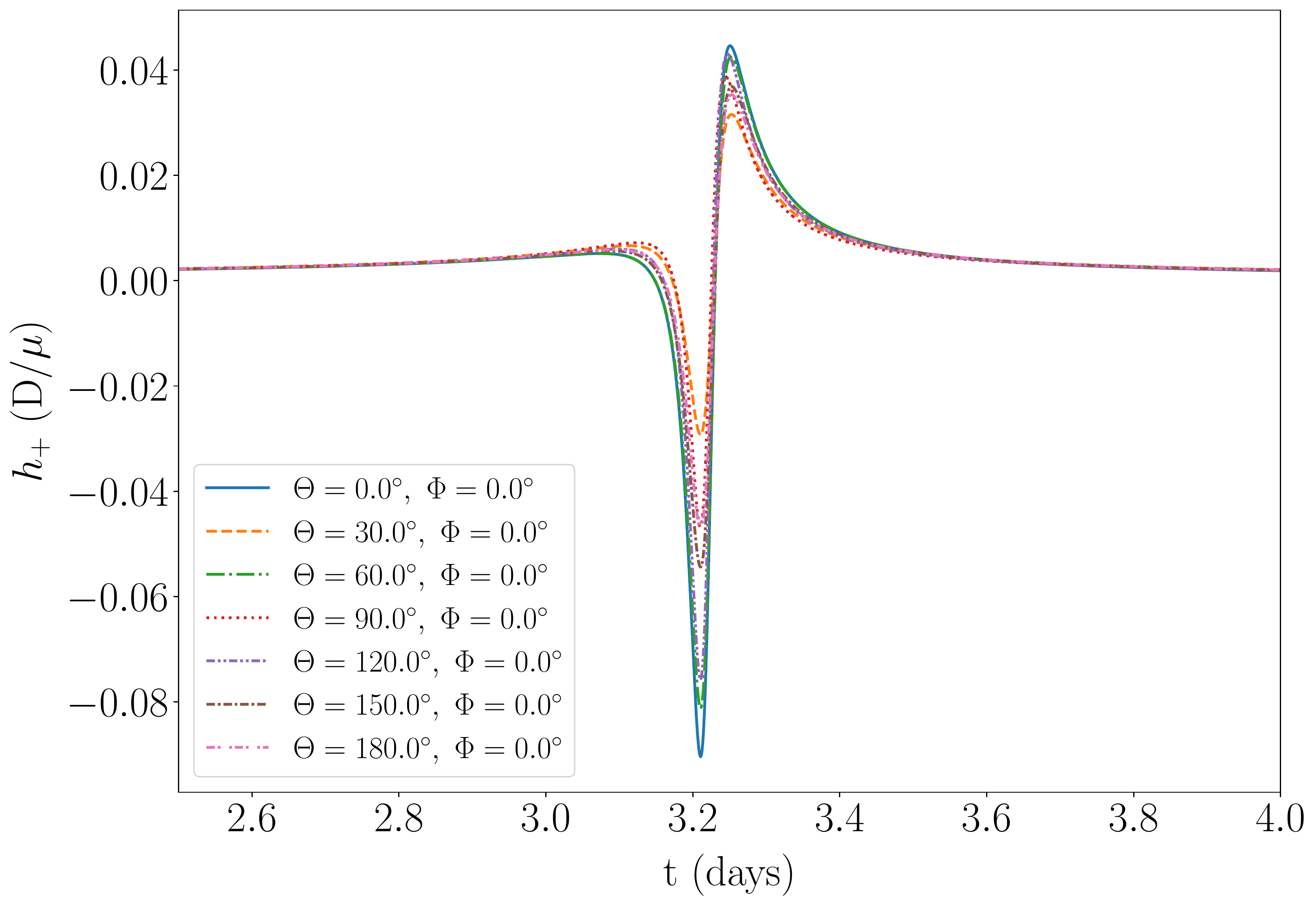} 
    \caption{$h_+$ Waveform}
  \end{subfigure}
  \begin{subfigure}[b]{0.5\linewidth}
    \centering
    \includegraphics[width=\linewidth]{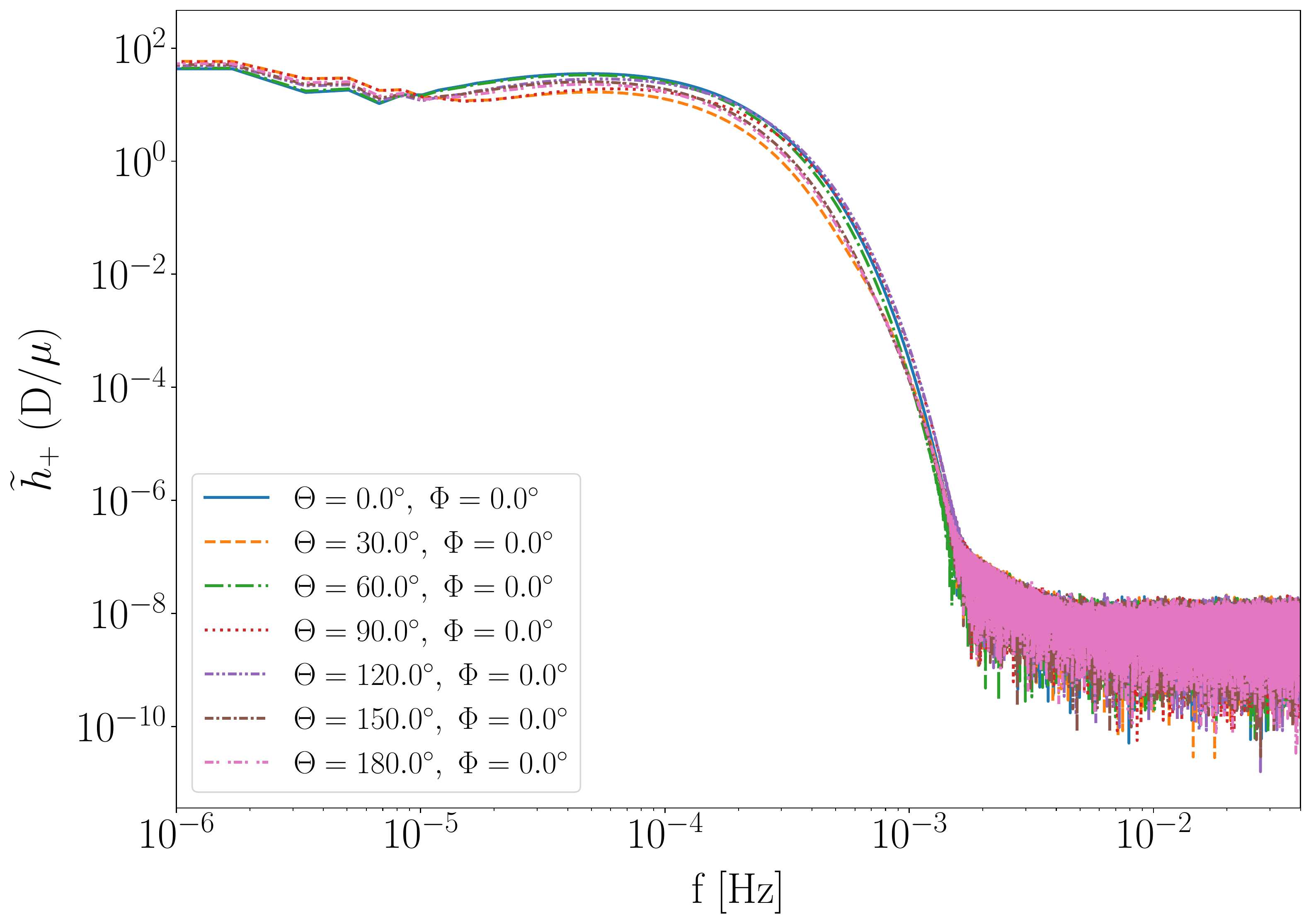}
    \caption{$h_+$ Spectra}
  \end{subfigure} 
  \begin{subfigure}[b]{0.5\linewidth}
    \centering
    \includegraphics[width=\linewidth]{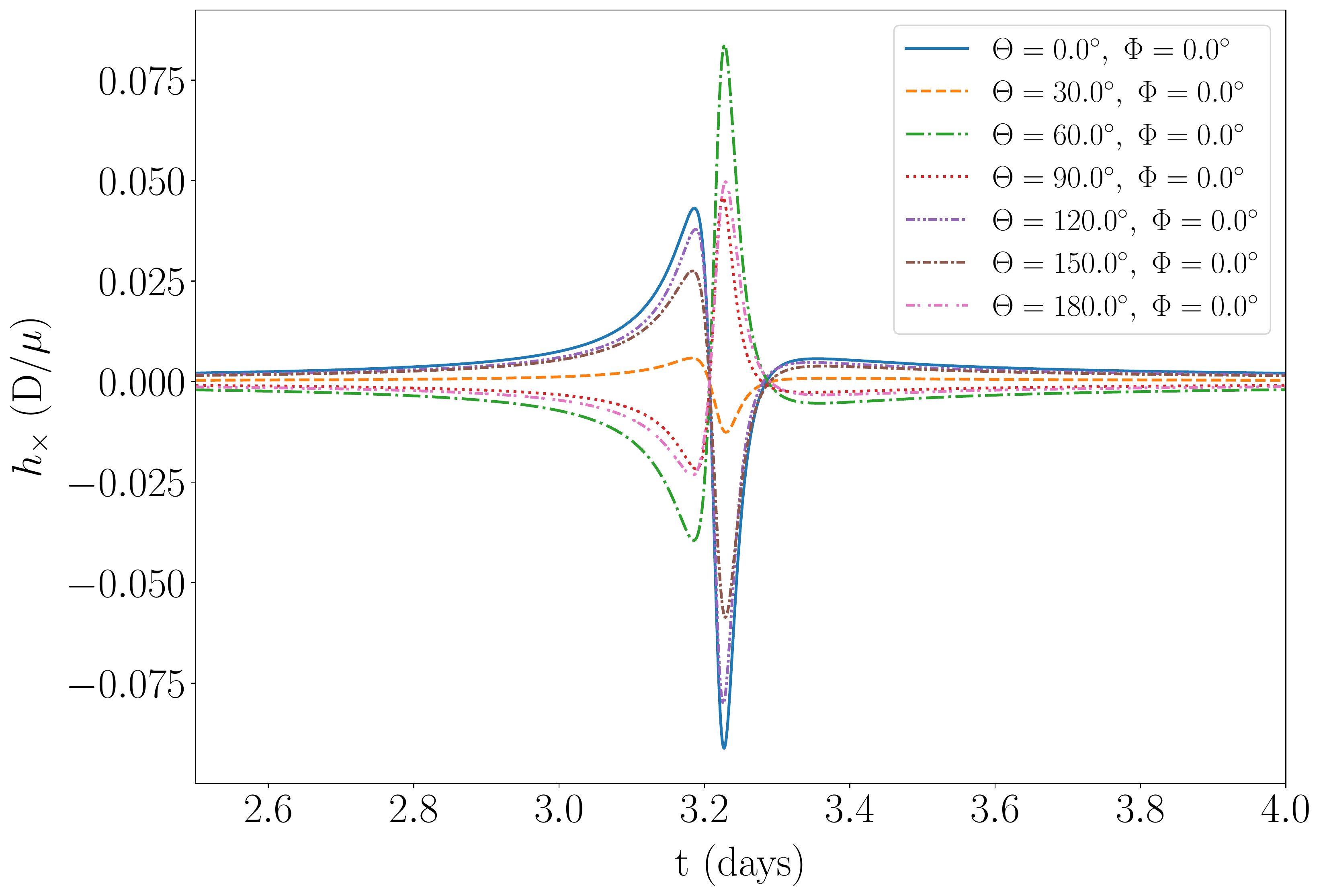} 
    \caption{$h_\times$ Waveform}
  \end{subfigure}
  \begin{subfigure}[b]{0.5\linewidth}
    \centering
    \includegraphics[width=\linewidth]{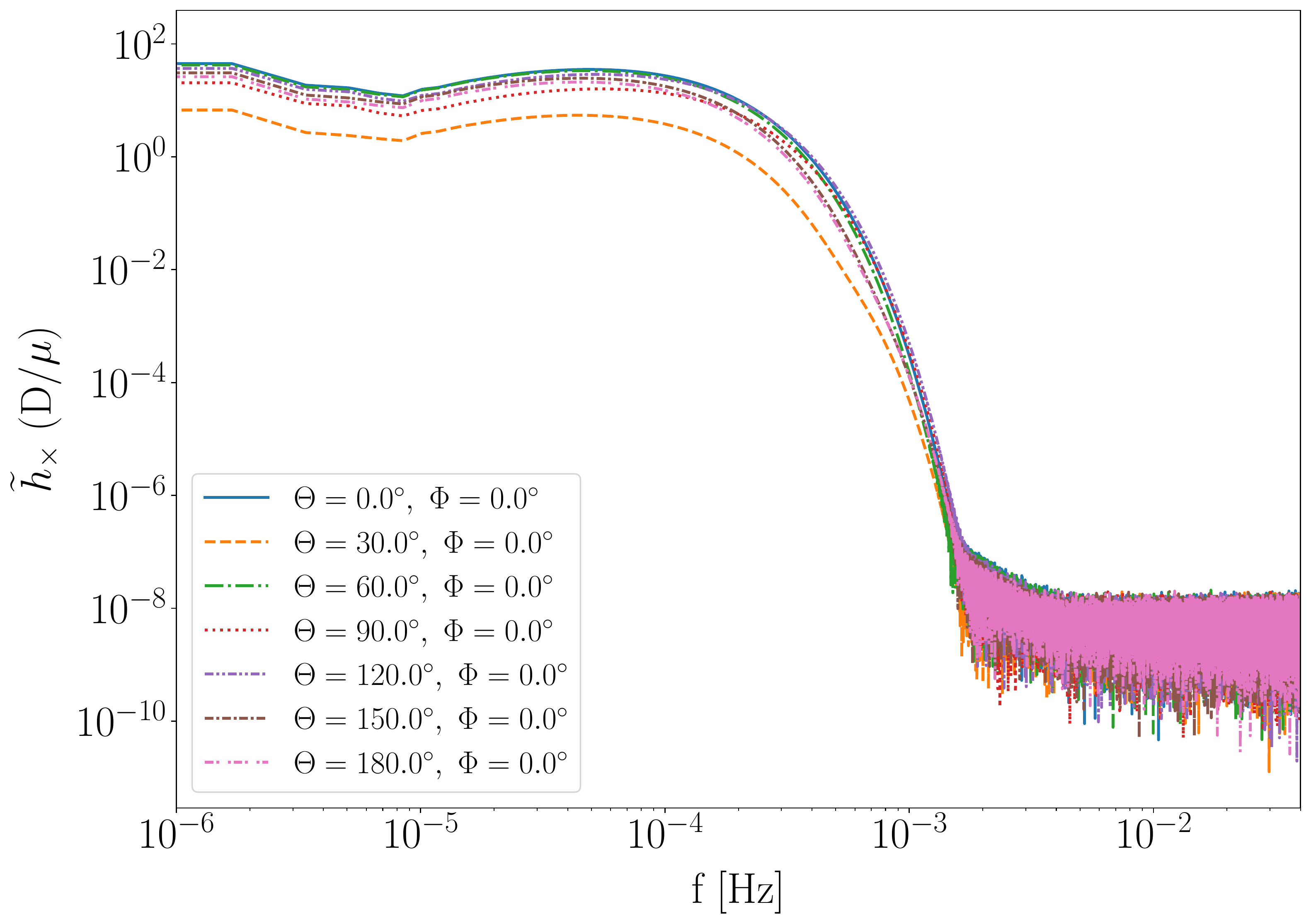}
    \caption{$h_\times$ Spectra}
  \end{subfigure} 
  \caption{Numerical Kludge waveform created with $a=0.9M$, $p=120M$, $e=0.999999$, $\iota=0.0^{\circ}$, $\Phi=0^{\circ}$ (face on) while adjusting the latitude viewing angle $\Theta$ from $0^\circ$ to $180^\circ$. In (a) and (c), we are showing the waveforms $h_+$ and $h_\times$ waveforms respectively. As opposed to the phase shift created by adjusting $\Phi$ in Figs. \ref{theta0} and \ref{theta90}, adjusting the latitude viewing angle $\Theta$ provides a much more significant change in the waveforms even between the $h_+$ and $h_\times$. $h_+$ waveform remains fairly constant in phase with a change in amplitude for the different angles. The $h_\times$ appears to have a different shape overall with a very noticeable $\Theta=30^\circ$ which has a much smaller amplitude compared to the other angles. This change is noticed in (b) and (d) which show the corresponding FFT spectra with a Tukey window, where you can see that the signals overlap very well for the $h_+$ and the $h_\times$ except for the $\Theta=30^\circ$ case for $h_\times$ where the amplitude is smaller. This deviation is important and implies that adjusting the latitude angle needs to be considered more carefully than the azimuthal viewing angle.}
   \label{phi0} 
\end{figure*}

\begin{figure*}
  {\large Adjusting Viewing Angle $\Theta$ for $a=0.9M$, $p=120M$, $e=0.999999$, $\iota=0^{\circ}$, and $\Phi=90^{\circ}$}\par\medskip
  \begin{subfigure}[b]{0.5\linewidth}
    \centering
    \includegraphics[width=\linewidth]{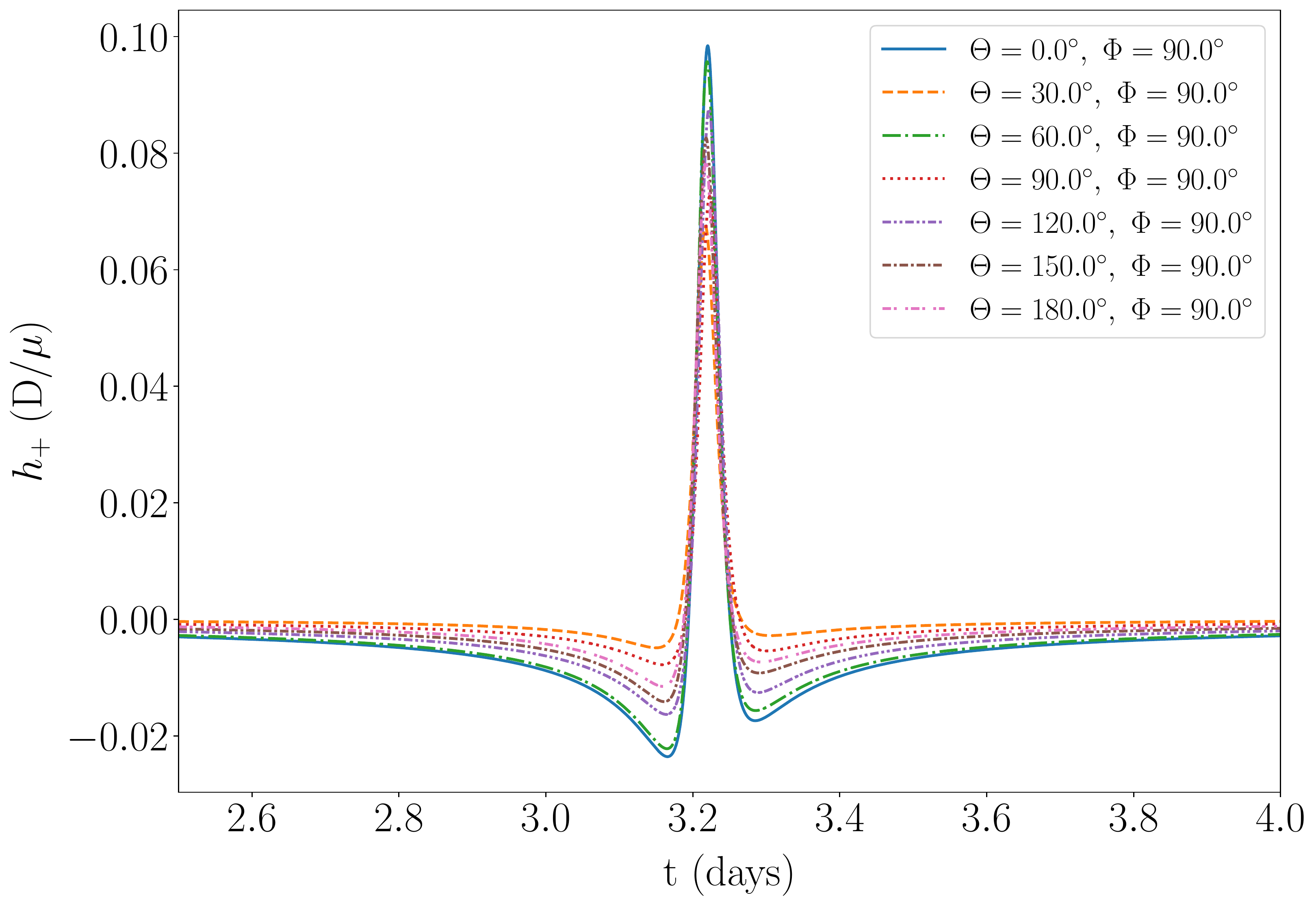} 
    \caption{$h_+$ Waveform}
  \end{subfigure}
  \begin{subfigure}[b]{0.5\linewidth}
    \centering
    \includegraphics[width=\linewidth]{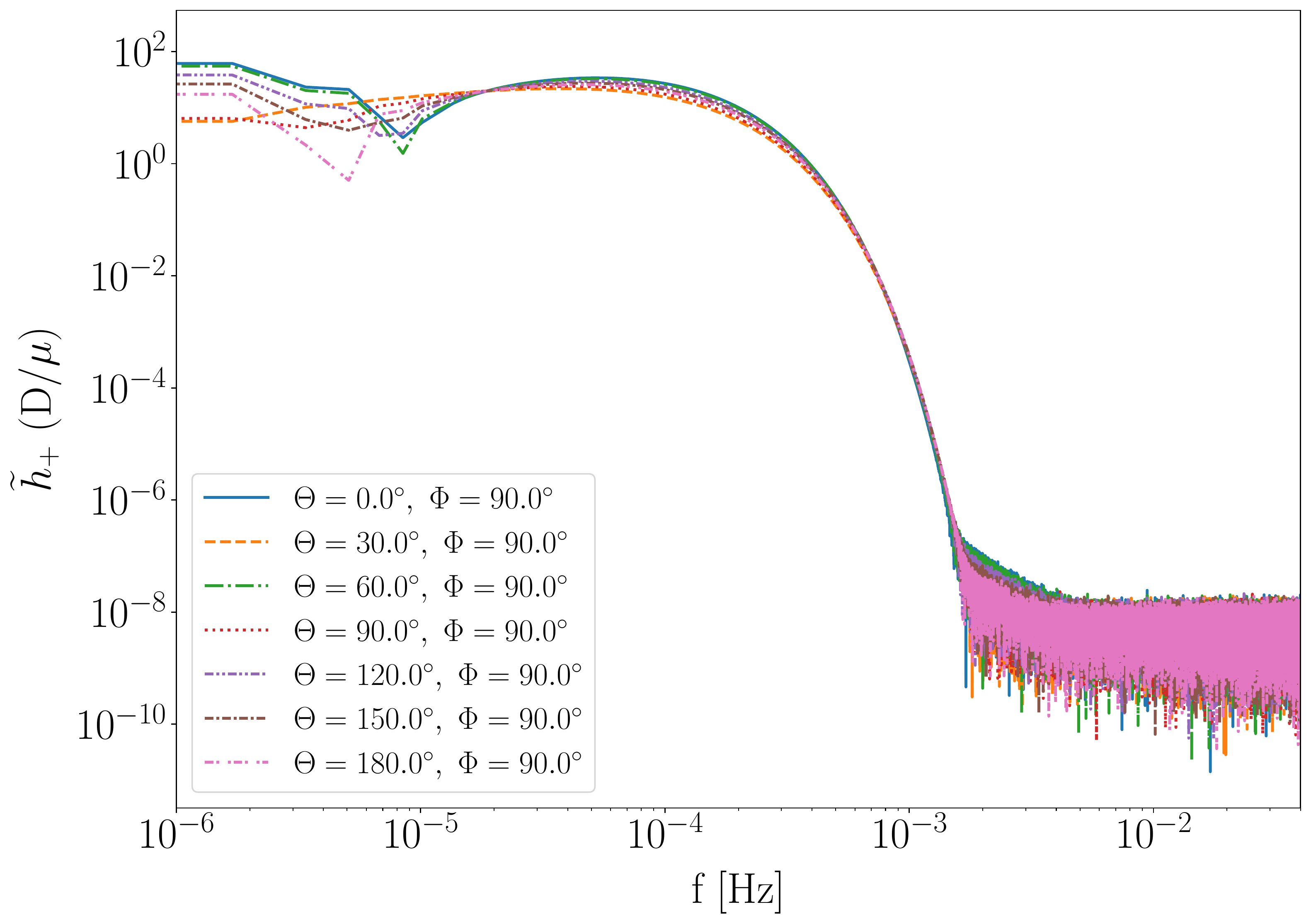}
    \caption{$h_+$ Spectra}
  \end{subfigure} 
  \begin{subfigure}[b]{0.5\linewidth}
    \centering
    \includegraphics[width=\linewidth]{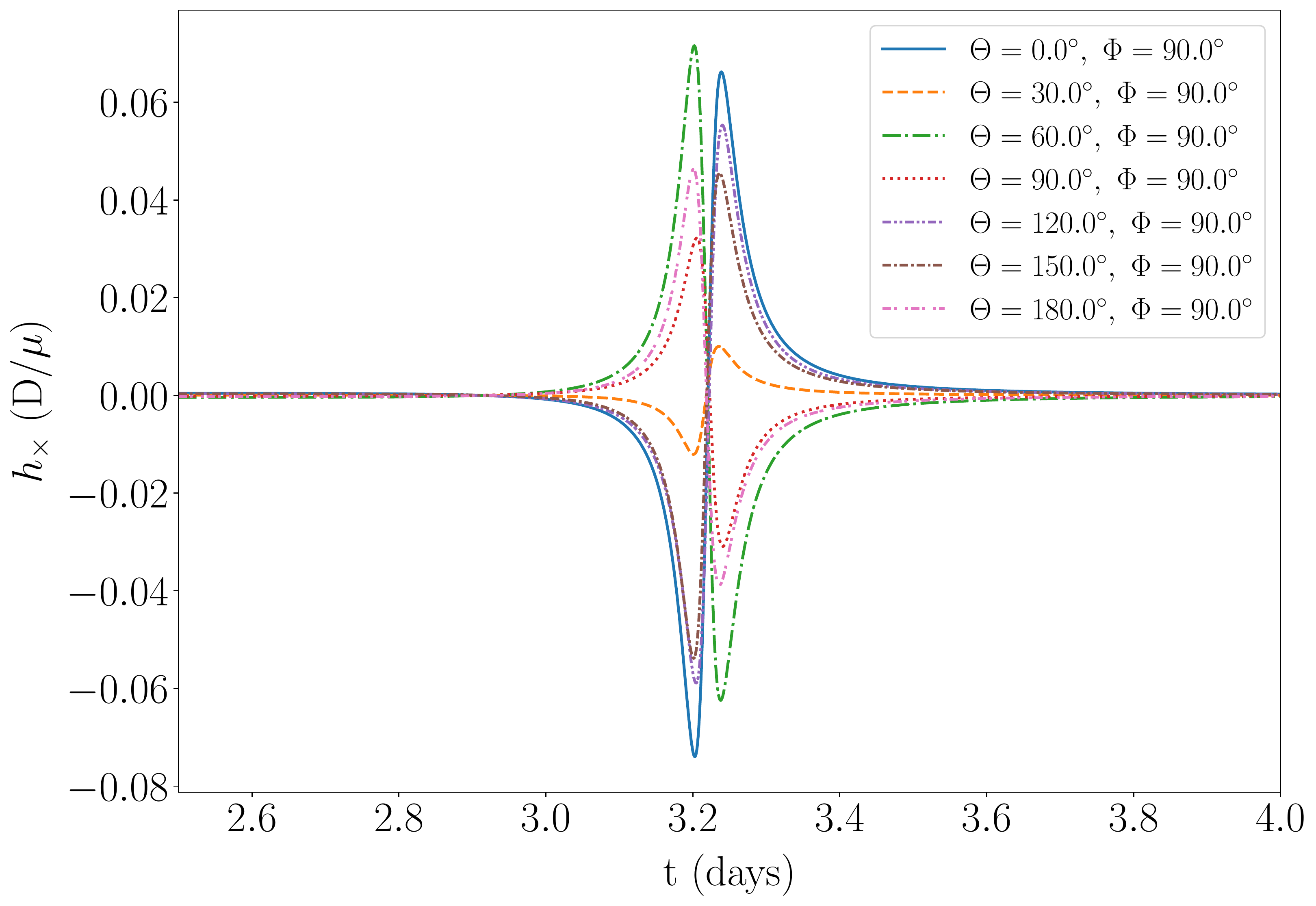} 
    \caption{$h_\times$ Waveform}
  \end{subfigure}
  \begin{subfigure}[b]{0.5\linewidth}
    \centering
    \includegraphics[width=\linewidth]{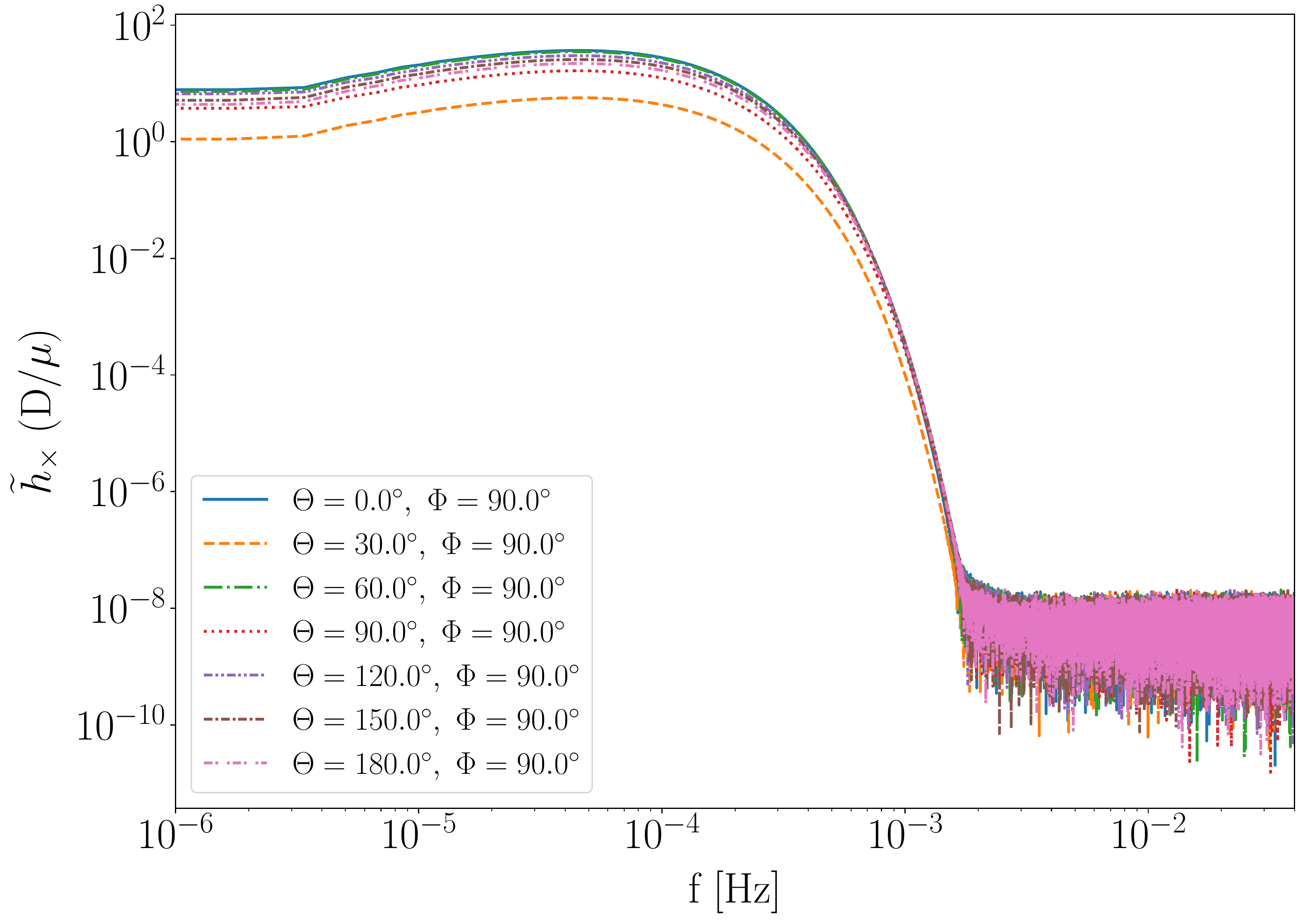}
    \caption{$h_\times$ Spectra}
  \end{subfigure} 
  \caption{Numerical Kludge waveform created with $a=0.9M$, $p=120M$, $e=0.999999$, $\iota=0.0^{\circ}$, $\Phi=90^{\circ}$ (edge on) while adjusting the latitude viewing angle $\Theta$ from $0^\circ$ to $180^\circ$. In (a) and (c), we are showing the waveforms $h_+$ and $h_\times$ waveforms respectively which as in Figure \ref{phi0}, the difference in the $h_+$ case seems to be a change in amplitude, while in the $h_\times$ case, there is an amplitude change as well as an overall different shape. Again for this case, $\Theta=30^\circ$ has a much smaller amplitude than the other angles, which is again noticeable in (b) and (d) which show the corresponding FFT spectra with a Tukey window. The $h_+$ spectra has a much more discernable change in the lower frequencies compared to Figure \ref{phi0}.}
   \label{phi90} 
\end{figure*}

While we have shown one iteration of the gravitational wave peep in Figure \ref{fig:waveform}, the actual shape, amplitude, and phase of the waveform can and will be adjusted based on the viewing angles of the observer $\{\Theta, \Phi\}$, the latitude and azimuthal angles respectively. Figs. \ref{theta0}-\ref{phi90} show the $h_+$ and $h_\times$ polarization waveforms and corresponding FFT spectra for different viewing angles. For Figs. \ref{theta0}, \ref{theta90}, we show a fixed latitude viewing angle {$\Theta$} at $0^{\circ}$ and $90^{\circ}$ respectively and then adjust the azimuthal viewing angle from $0^{\circ}$ to $180^{\circ}$, where $0^{\circ}$ is the face-on-case. For Figure \ref{theta0}, the spectra in the LISA bandwidth is very similar with significant overlap, it is only in the much lower frequencies that there is a noticeable deviation. Figure \ref{theta90} shows the $\Theta=90^{\circ}$ case and the waveforms have a much lower amplitude overall compared to Figure \ref{theta0}, however, the spectra have very similar amplitudes. The spectra for Figure \ref{theta0} also align very similarly throughout most of the LISA bandwidth with a larger separation in the lower frequencies. This implies that the azimuthal observing angle of the gravitational wave peep will not significantly affect the final spectrum in the LISA bandwidth. \par

Figs. \ref{phi0}, \ref{phi90} are very similar, though here, the azimuthal angle {$\Phi$} is kept fixed at $0^{\circ}$ and $90^{\circ}$ respectively. The difference between adjusting $\Theta$ versus $\Phi$ is that the latitude angle changes the overall shape of the waveform more so than the azimuthal angle which just acts as a phase shift. The phase shift makes sense because if the system is rotated with respect to the observer then the gravitational radiation will be emitted at a slightly different time. Overall between Figs. \ref{phi0} and \ref{phi90}, the spectra seem to be more differentiated compared to the change in $\Phi$ which indicates that adjusting $\Theta$ seems to have a larger impact on the resulting signal which will need to be accounted for. This separation is best demonstrated by the $h_\times$ cases where noticeably the $\Theta=30^\circ$ case has a much smaller amplitude in the waveforms and the spectra for Figs. \ref{phi0} and \ref{phi90}.\par

\section{Conclusion}

It is well known that the later stages of an EMRI inspiral are characterized by moderate to low eccentricities because of the circularization tendency of radiation damping. For this reason, and because of the inherent difficulty of calculating waveforms from highly eccentric orbits, the early stages of the inspiral have received less attention. Using the numerical kludge approach, which permits quite accurate modeling of highly eccentric orbits, we present here (in Figure \ref{inspiral}) a complete inspiral waveform, albeit at a fairly low resolution (with larger time steps to accommodate the length of signal). Nevertheless, the resolution is sufficient to catch the ``peep” of the signal, which is a relatively brief part of each orbit. During the peep, the signal effectively switches from a long period, small amplitude waveform into one which is in the LISA bandwidth and with a far greater amplitude. The inset in Figure \ref{inspiral} is a much higher resolution snapshot of one of the peeps, giving a close-up of just a tiny portion of the early inspiral orbit.\par

If we look for evolution in the very long waveform, over the course of 600 million years, we know that the period of each orbit reduces steadily, and the frequency of the peep itself also increases, so the ``chirp” characteristic of binary inspirals is observed (though difficult to see at the scale of Figure \ref{inspiral}). However one’s eye does notice that there is surprisingly little amplitude evolution, the other “chirp” characteristic, for most of the inspiral. This is because it is only the amplitude of the peep which is discernable in Figure \ref{inspiral}. The main orbit waveform is simply too small in amplitude to be noticeable. Essentially Figure \ref{inspiral} represents the ``chirp” of the ``peep”, and the peep does not do as much “chirping” as we might expect, at least in terms of amplitude increase. The reason is probably that a highly eccentric orbit principally evolves by reducing its apoapsis. The periapsis radius ($r_p$) barely changes from one orbit to the next, and thus the amplitude evolves relatively little. Only at the tail end of the inspiral, when the eccentricity becomes moderate, do we see significant amplitude evolution in Figure \ref{inspiral}.\par

Unfortunately, a single peep does not likely represent a detectable signal for LISA, because they are repeated so rarely, given the long-period nature of the typical peep orbit. However, at any given moment there must be a large number of peeps being emitted across the cosmos, and modeling of this confusion noise signal is definitely a challenge for LISA data analysis.  The choice of the term ``peep” is intended to evoke the small sound of a chick peeping away. While an individual may not be particularly loud, an ensemble may create enough background noise to drown out conversation. While the outline of the EMRI signal confusion noise has been reasonably accurately laid down in the work of Barack and Cutler \cite{barack_confusion_2004}, Bonetti and Sesana \cite{bonetti_gravitational_2020}, and EMRB signals with Toonen et al. \cite{toonenGravitationalWaveBackground2009} and Fan et al. \cite{fanExtrememassratioBurstDetection2022} we propose to model it more precisely using the basic approach discussed in this paper, the numerical kludge applied to complete orbits with sufficient resolution to catch the individual peeps and include them in the full spectrum for LISA.\par

We have shown the effects of different window functions as well as how changing each of the orbital parameters adjusts the resulting FFT spectra. The window functions show that the Hann, Blackman, Tukey, and Nuttall windows seem most appropriate for this study. For signals with several complete orbits, the Nuttall window performs especially well. Figs. \ref{fig:SpecCompA}-\ref{fig:SpecCompIota} show the effects of changing each of the orbital parameters and indicate that the MBH spin and CO orbital eccentricity at this stage will not impact the spectra significantly on their own; however, the semi-latus rectum and orbital inclination do seem to impact the results much more noticeably. While altering the orbital eccentricity doesn't result in significant differences between a high eccentricity (such as $e=0.99$) and an eccentricity which is arbitrarily close to parabolic (such as $e=0.999999999$ where in one year of data there is a single GW emission) in spectra there does seem to be a noticeable difference as indicated in Figure \ref{fig:Parabolic} which shows a significant amplitude deviation and an overall larger width of the signal in frequency for the ``peep" compared to the parabolic burst. This suggests that approximating peeps as parabolic incidences may not be especially accurate. We also found that adjusting the azimuthal viewing angle does not significantly impact the resulting spectra; however, the latitude viewing angle does seem to result in a slight deviation at certain values.\par

To determine the potential for resolving individual sources from the background noise, we will use our study of the gravitational wave peep. Then using relativistic population data and an updated black hole mass function from the Illustris Project we can expand our project \cite{nelson_illustris_2015,sijacki_illustris_2015,vogelsberger_properties_2014}. In a future paper, we will combine the parameter estimations discussed in this paper, population data from the updated black hole mass function using Illustris, and a reasonable estimate for the rate of EMRIs for a given MBH mass to generate an estimate for the number of captures as a function of redshift in an ensemble of galaxies. We will use this to estimate the gravitational wave background from peeps. To this end, our study will provide an update on the EMRI signal confusion noise problem.

\section{Acknowledgements}

We would like to thank Ben Bogner, Calla Bassett, and Harry O'Mara for their assistance in the early stages of this project as well as Woodrow Gilbertson, Paul Bonney, Erik Monson, and Jonathan Thompson for their helpful discussion.\par 
The data that support the findings of this study are openly available on Zenodo \cite{oliver_2024}. \par
This research is supported by the Arkansas High Performance Computing Center which is funded through multiple National Science Foundation grants and the Arkansas Economic Development Commission.\par
ADJ acknowledges support from the Caltech and Jet Propulsion Laboratory President’s and Director’s Fund. \par
KG acknowledges support from research grant PID2020-1149GB-I00 of the Spanish Ministerio de Ciencia e Innovaci{\'o}n.\par
This paper utilized \texttt{SciPy} \cite{scipy}, \texttt{NumPy} \cite{numpy}, \texttt{Matplotlib} \cite{matplotlib}, Mathematica \cite{Mathematica}, and the \texttt{Black Hole Perturbation Toolkit} \cite{BHPToolkit}.

\clearpage

\bibliography{iopart-num}

\end{document}